\documentclass[trackchanges]{aastex701}
\shorttitle{jet property and accretion regime for blazar sequence}
\shortauthors{Q.C. Long}

\usepackage{longtable}
\usepackage{amsmath}
\usepackage{array}
\usepackage{hyperref} 
\usepackage{xcolor}
\usepackage{graphicx}
\usepackage{textcomp}

\begin{document}

\title{The Jet Properties and Accretion Regime for the Blazar Sequence}

\correspondingauthor{Qingchen. Long}

\author[orcid=0009-0003-4213-7662]{Qingchen. Long}
\affiliation{Center for Astrophysics, Guangzhou University, Guangzhou 510006, People’s Republic of China; qclong@e.gzhu.edu.cn}
\affiliation{Greater Bay Brand Center of the National Astronomical Data Center, Guangzhou 510006, People’s Republic of China}
\affiliation{Astronomy Science and Technology Research Laboratory of Department of Education of Guangdong Province, Guangzhou 510006, People’s Republic of China}
\email[show]{qclong@e.gzhu.edu.cn}

\begin{abstract}

The puzzling bimodality displayed by blazars on the broad-band spectral indices ($\alpha_{\rm ro}-\alpha_{\rm ox}$) plane is still an open question. To investigate its physical origin, we compiled a sample comprising 136 flat-spectrum radio quasars (FSRQs), 64 low-synchrotron-peaked BL Lac objects (LBLs), 39 intermediate-synchrotron-peaked BL Lac objects (IBLs), and 105 high-synchrotron-peaked BL Lac objects (HBLs). Our results show that (FSRQs+LBLs)$\rightarrow$IBLs$\rightarrow$HBLs follow a \reflectbox{$\angle$}-shaped evolutionary sequence on the $\alpha_{\rm ro}-\alpha_{\rm ox}$ plane, with the average 5\,GHz radio Doppler factor, jet power, and accretion Eddington ratio decreasing along this \reflectbox{$\angle$}-shaped track. Additionally, the distributions of $\alpha_{\rm ox}$ exhibit only slight differences among blazar subclasses, whereas those of $\alpha_{\rm ro}$ and $\alpha_{\rm rx}$ differ significantly. Our results can be naturally explained if the distribution of blazars on the $\alpha_{\rm ro}-\alpha_{\rm ox}$ plane is regulated by the accretion rates and the Doppler beaming effect across different bands ($\delta_i^{q_i}/\delta_j^{q_j}$). However, $\delta_i/\delta_j$ plays a pivotal role in the distribution, while jet shape ($q=2(3)+\alpha_\nu$) and spectral indices ($\alpha_\nu$) play a minor role. Our results suggest that FSRQs+LBLs are strong-jet sources coupled with radiatively efficient accretion, and their jets are accelerated to a larger distance from the core and still maintain relativistic speeds down to the radio region; HBLs are weak-jet sources coupled with radiatively inefficient accretion, and their jets may begin to decelerate in the optical region and exhibit sub-relativistic speeds as they propagate to the radio region. IBLs are regarded as transitional objects between FSRQs+LBLs and HBLs.

\end{abstract}



\section{Introduction}\label{Sec-intro}

Blazars are a subclass of active galactic nuclei (AGNs) with a relativistic jet aligned with our line of sight \citep{Urry&Padovani1995PASP}, which causes the broad-band emissions of blazars to be boosted and confers them with many extreme properties \citep[e.g., super-luminal motions, rapid variability, high and variable polarization, high luminosity, and high-energy $\gamma$-ray emissions; see][]{Wills1992ApJ,Fan1996A&A,Fan2000ApJ,Fan2005ChJAS,Fan2013RAA,Raiteri2017Natur,Abdollahi2022ApJS}. Blazars consist of two subclasses, namely flat-spectrum radio quasars (FSRQs) and BL Lac objects (BL Lacs), distinguished by the equivalent width (EW) of their optical emission lines. FSRQs exhibit strong broad emission lines with $\rm EW>5\AA$, while BL Lacs have no or weak emission lines characterized by $\rm EW<5\AA$ \citep{Scarpa1997A&A}. The discrepancy of emission lines between FSRQs and BL Lacs was explained by the different accretion regimes or the Doppler beaming effect \citep[e.g.,][]{Corbett2000MN,Fan2003ApJL,Ghisellini2008MN,Ghisellini2011MN,Padovani2012MN,Foschini2012RAA,Sbarrato2014MN,Chen2025MN}.

The typical feature of blazars is a double-hump structure in the spectral energy distribution \citep[SED; see][]{Fossati1998MN,Donato2001A&A,Fan2016ApJS,Blandford2019,Yang2022ApJS}. The lower-energy hump, which spans from the radio to X-ray bands, is characterized by synchrotron radiation; the higher-energy hump is located at the MeV to TeV bands \citep{Fossati1998MN,Nieppola2006A&A,Fan2016ApJS}. However, the origin of the high-energy bump remains controversial. It may arise either from inverse-Compton scattering of soft photons by relativistic electrons (the leptonic model) or from synchrotron emission by ultrarelativistic protons and hadronic cascades (the hadronic model; see \citealt{Bottcher2013ApJ,Cerruti2015MN}). Based on the location of the peak frequency ($\nu_{\rm p}^{\rm syn}$) of the synchrotron hump in the SED, BL Lacs are divided into low-synchrotron-peaked BL Lacs (LBLs: $\nu_{\rm p}^{\rm syn}<10^{14}$\,Hz), intermediate-synchrotron-peaked BL Lacs (IBLs: $\rm 10^{14}\,Hz<\nu_p^{syn}<10^{15}\,Hz$), and high-synchrotron-peaked BL Lacs \citep[HBLs: $\nu_{\rm p}^{\rm syn}>10^{15}$\,Hz; see][]{Abdo2010ApJ}. FSRQs are usually powerful sources with $\nu_{\rm p}^{\rm syn}\lesssim10^{14}$\,Hz \citep{Abdo2010ApJ,Giommi2012A&A,Fan2016ApJS,Yang2022ApJS}. BL Lacs are classified as radio-selected BL Lacs (RBLs) and X-ray-selected BL Lacs (XBLs) based on the surveys that discovered them; RBLs tend to be LBLs, and XBLs tend to be HBLs \citep{Giommi1995A&AS,Wu2007A&A}. Additionally, BL Lacs are also divided into LBLs and HBLs according to the broad-band spectral index (also known as the effective spectral index) from the 5\,GHz radio band to the 1\,keV X-ray band \citep[$\alpha_{\rm rx}$, the broad-band spectral indices $\alpha_{ij}$ are defined in the following as Eq\,(\ref{Eq-alpha}); see][]{Ledden1985ApJ}. LBLs have $\alpha_{\rm rx}>0.75$, while HBLs have $\alpha_{\rm rx}<0.75$ \citep{Padovani1995ApJ,Bai2001ApJ}.

The above-mentioned differences in observational properties among blazar subclasses suggest that the underlying physics differs among them. The differences in SEDs likely suggest different accretion regimes \citep{Bottcher2002ApJ,Wang2002ApJ,Ghisellini2008MN,Paliya2021ApJS}. \cite{Keenan2021MN} suggested that FSRQs, LBLs, and IBLs are strong-jet sources associated with radiatively efficient accretion, while HBLs are weak-jet sources that exhibit radiatively inefficient advection-dominated accretion flows \citep[ADAFs, see also][]{Meyer2011ApJ,Ye2025A&A}. Furthermore, differences in the relativistic jet effect among blazars have also been found, namely, an anti-correlation between the radio Doppler factor ($\delta_{\rm r}$) and $\nu_{\rm p}^{\rm syn}$ \citep[e.g.,][]{Wu2007A&A,Nieppola2008A&A,Yang2022ApJ,Long2025ApJ}, which implies different jet properties among the blazar subclasses. Interestingly, Very Long Baseline Array (VLBA) observations on parsec scales show that the radio knots in FSRQs exhibit super-luminal motions, LBLs exhibit sub- to super-luminal motions, whereas the radio components in HBLs generally move slowly and, in some cases, remain stationary \citep[e.g.,][]{Jorstad2001ApJS,Jorstad2017ApJ,Piner2008ApJ,Piner2010ApJ,Lister2009AJ,Lister2013AJ,Thevenet2026A&A}. Obviously, VLBA observations are consistent with the anti-correlation of $\delta_{\rm r}-\nu_{\rm p}^{\rm syn}$. Observational studies across different blazar subclasses seem to support an emerging consensus that strong-jet sources are associated with accelerating jets, whereas weak-jet sources are associated with decelerating jets \citep[e.g.,][]{Fan1994Ap&SS,Georganopoulos2003ApJ,Ghisellini2005A&A,Piner2008ApJ,Piner2010ApJ,Karamanavis2016A&A,Jorstad2017ApJ,Liodakis2018ApJ,Long2025ApJ,Long2026MN}.

A crucial way of understanding the blazar physics is to study their SED behavior, which can be parameterized using the broad-band spectral indices \citep{Balmaverde2006A&A}. \cite{Padovani1995ApJ} found that LBLs and HBLs dwell in two completely different regions of the $\alpha_{\rm ro}-\alpha_{\rm ox}$ plane \citep[see also][where the broad-band frequencies are usually defined at the 5\,GHz radio band, the optical $V$ band, and the 1\,keV X-ray band]{Ledden1985ApJ,Donato2001A&A,Balmaverde2006A&A,Nieppola2006A&A,Abdo2010ApJ,Fan2016ApJS,Yang2022ApJS}. Most LBLs occupy the upper right region (URR) of the $\alpha_{\rm ro}-\alpha_{\rm ox}$ plane with $\alpha_{\rm rx}>0.75$, most HBLs occupy the lower left corner of the $\alpha_{\rm ro}-\alpha_{\rm ox}$ plane with $\alpha_{\rm rx}<0.75$, and IBLs appear to serve as a bridge between LBLs and HBLs \citep{Nieppola2006A&A}. Interestingly, FSRQs occupy the region of LBLs on the $\alpha_{\rm ro}-\alpha_{\rm ox}$ plane \citep{Donato2001A&A,Abdo2010ApJ,Fan2016ApJS,Yang2022ApJS}.

The puzzling bimodality displayed by blazars on the $\alpha_{\rm ro}-\alpha_{\rm ox}$ plane indicates their different energy distributions in radio, optical, and X-ray bands, which was explained as the result of migration of $\nu_{\rm p}^{\rm syn}$ \citep[see Figure\,12 of][]{Padovani1995ApJ}. In the unified model of AGNs, the discrepancy of observational properties is regulated by accretion and orientation \citep{Shen2014Nat}. However, the jet orientation makes it difficult for any observational properties of blazars to avoid the impact of Doppler beaming effect. Therefore, the observational properties of blazars should be related to the accretion and the beaming effect. Several early studies have suggested that the beaming effect could affect the distribution of sources on the $\alpha_{\rm ro}-\alpha_{\rm ox}$ plane \citep[see][]{Fan1996A&A,Chiaberge2000A&A,Capetti2000MN,Trussoni2003A&A,Balmaverde2006A&A}. However, this proposal remains a theoretical prediction, and direct observational evidence has yet to be obtained. In this paper, we aim to (1) verify whether the beaming effect impacts the distribution of blazars on the $\alpha_{\rm ro}-\alpha_{\rm ox}$ plane and investigate how the beaming effect influences the distribution of blazars, (2) explore accretions and jet physics behind this phenomenological bimodality.

The structure of the paper is as follows. In Section\,\ref{Ses-sample}, we describe our blazar sample in detail, calculate the broad-band spectral indices and some parameters. In Section\,\ref{Sec-result}, we present our results. The discussions and summaries of our results are presented in Sections\,\ref{Sec-discu} and \ref{Sec-summary}, respectively. Throughout this paper, we adopt a flat-$\rm \Lambda CDM$ cosmology with $H_0=70\,\rm{km\,s^{-1}\,Mpc^{-1}}$, $\Omega_\Lambda=0.73$, $\Omega_{\rm M}=0.27$, and the web-based cosmology calculator \citep{Wright2006PASP}.

\section{Blazar Sample}\label{Ses-sample}

To achieve the objectives of this work, our sample includes all subclasses of blazars. In addition, the selected blazars were required to satisfy the following criteria: (1) available measurements of $\nu_{\rm p}^{\rm syn}$, allowing us to classify the sources; (2) available radio, optical, and X-ray fluxes, allowing us to calculate the broad-band spectral indices; (3) available Doppler factors ($\delta_\nu$), allowing us to investigate whether Doppler beaming affects the distribution of blazars on the $\alpha_{\rm ro}-\alpha_{\rm ox}$ plane; (4) available low-frequency radio fluxes, allowing us to estimate the intrinsic jet power ($P_{\rm jet}^{\rm in}$); and (5) available accretion luminosity ($L_{\rm disk}$) and dynamical black hole (BH) mass ($M_{\rm BH}$), allowing us to estimate the accretion rates (which are quantified by the ratio of $L_{\rm disk}$ to the Eddington luminosity ($L_{\rm Edd}$): $\lambda_{\rm Edd}=\log(L_{\rm disk}/L_{\rm Edd})$). However, most BL Lacs have no available $L_{\rm disk}$ or $M_{\rm BH}$. To obtain as large as possible a blazar sample, blazars that meet the first four criteria are included.

The third data release of the Fourth Catalog of AGNs Detected by $Fermi$-LAT \citep[4LAC-DR3;][]{Ajello2022ApJS} provides $\nu_{\rm p}^{\rm syn}$ measurements for sources in the 4FGL catalog (although a few blazars do not have available $\nu_{\rm p}^{\rm syn}$ in 4LAC-DR3, we find that their $\nu_{\rm p}^{\rm syn}$ are available in \citealt{Nieppola2006A&A,Xiong2015MN-a,Chang2019A&A}). Several studies have estimated the $\delta_{\rm r}$ for blazars \citep[$\delta_{\rm r}$-catalogs; see][]{Hovatta2009,Wu2014,Liodakis2017MN,Liodakis2018ApJ,Ye2021PASJ}. We therefore cross-matched 4LAC-DR3 with these $\delta_{\rm r}$-catalogs and further examined whether the matched sources have available broad-band data.
In total, we constructed a sample containing 136 FSRQs, 64 LBLs, 39 IBLs, and 105 HBLs that have available $\nu_{\rm p}^{\rm syn}$, 5 GHz core radio flux{\color{red},} optical flux, X-ray flux, $\delta_{\rm r}$, and low-frequency radio flux. Most of our blazars (281 blazars; see Table\,\ref{table3}) have available $M_{\rm BH}$.

\subsection{The Collection and Analysis of Data}\label{Sec-2.1}

Most of the $\nu_{\rm p}^{\rm syn}$ of our blazars are taken from \cite{Ajello2022ApJS}, only a few from \cite{Nieppola2006A&A}, \cite{Xiong2015MN-b}, and \cite{Chang2019A&A}.

To mitigate the possible contamination from the extended emissions \citep{Panessa2019NatAs}, the 5 GHz core radio flux ($f_{\rm r}$) is adopted in our analysis. The available $f_{\rm r}$ are extracted from the NASA/IPAC Extragalactic Database (NED\footnote{https://ned.ipac.caltech.edu/}) and the literatures (see table\,\ref{table1}). In this work, the optical emissions at the $5.4143\times10^{14}$\,Hz $V$ band are adopted; the observational optical fluxes of our blazars are taken from the SIMBAD database, operated at CDS Strasbourg, France (CDS)\footnote{https://simbad.u-strasbg.fr/simbad/}. In line with previous works \citep{Ledden1985ApJ,Fan2016ApJS}, we use the observational 1\,keV X-ray flux ($f_{\rm 1keV}$) for our analysis. However, only a few blazars have available $f_{\rm 1keV}$. Therefore, for most of our blazars, $f_{\rm 1keV}$ are extrapolated from other approximate bands (e.g., $0.1-2.4$\,keV and $0.3-2$\,keV) using $f_\nu \propto \nu^{-\alpha_{\rm x}}$ ($\alpha_{\rm x}=\Gamma_{\rm x}-1$, where $\Gamma_{\rm x}$ is the power-law photon index). The typical values of $\Gamma_{\rm x}$, taken from \cite{Donato2005A&A}, are adopted as 1.59, 1.92, and 2.24 for FSRQs, LBLs+IBLs, and HBLs, respectively. The X-ray fluxes of our blazars are mainly obtained from NED, with a few taken from the literatures. The broad-band spectral indices ($\alpha_{\rm ro}$, $\alpha_{\rm ox}$, and $\alpha_{\rm rx}$) are calculated using the equation as follows \citep{Ledden1985ApJ}:

\begin{equation}
    \alpha_{ij}=-\log(f_i/f_j)/\log(\nu_i/\nu_j)
    \label{Eq-alpha}
\end{equation}
where $f_\nu$ and $\nu$ correspond to the flux and frequency, respectively.

The strength of jet is the most intuitive indicator of jet properties, and can be quantified as $P_{\rm jet}^{\rm in}$ \citep[e.g.,][]{Meyer2011ApJ,Keenan2021MN}. Usually, $P_{\rm jet}^{\rm in}$ is represented by the low-frequency radio luminosity ($L_{\rm r}^{\nu_{low}}$), as it is only marginally affected by Doppler-boosted compact components \citep{Wu2007A&A,Keenan2021MN}. The formula is as follows \citep[e.g.,][]{Cavagnolo2010ApJ}

\begin{equation}
    \log P_{\rm jet}^{\rm in}=0.64(\log L_{\rm r}^{\nu_{low}} -40)+43.54
    \label{Eq-(P-L)}
\end{equation}
In this work, we adopt the 340\,MHz radio fluxes ($f_{\rm 340MHz}$) to calculate $P_{\rm jet}^{\rm in}$, which are taken from CDS. For blazars without available $f_{\rm 340MHz}$, the other approximate wave bands (e.g., 150\,MHz, 330\,MHz, 352\,MHz, 365\,MHz, and 843\,MHz) are extrapolated to $f_{\rm 340MHz}$ using $f_\nu \propto \nu^{-\alpha_{\rm r}}$ \citep[$\alpha_{\rm r}=0.7$, e.g.,][]{Keenan2021MN}.

The Doppler beaming effect, quantified by $\delta_\nu$, is often considered in blazar physics. However, $\delta_\nu$ has generally been estimated in the radio and $\gamma$-ray bands \citep[e.g.,][]{Ghisellini1993ApJ,Hovatta2009,Wu2007A&A,Wu2014,Fan2014RAA,Liodakis2017MN,Liodakis2018ApJ,Pei2022ApJ}, whereas estimates are generally unavailable in the optical and X-ray bands. The $\delta_{\rm r}$ of our blazars are taken from \cite{Hovatta2009} (22\,GHz \& 37\,GHz), \cite{Wu2014} (5\,GHz), \cite{Liodakis2017MN} (see their table\,1), \cite{Liodakis2018ApJ} (15\,GHz), and \cite{Ye2021PASJ} (5\,GHz). In theory, $\delta_\nu$ is determined by the velocity ($\beta$) of plasma flows and the view angle ($\theta$) to our line of sight: $\delta_\nu=[\Gamma(1-\beta {\rm cos}\theta)]^{-1}$, where $\Gamma$ is the bulk Lorentz factor ($\Gamma=1/\sqrt{1-\beta^2}$). Considering the possible acceleration and deceleration of blazar jets, $\delta_\nu$ may vary among different emission bands. Because we adopt the 5\,GHz core radio emissions in this paper, $\delta_\nu$ at 5\,GHz band ($\delta_{\rm 5GHz}$) is adopted. Conservatively, we extrapolate $\delta_\nu$ at other bands to $\delta_{\rm 5GHz}$ using the empirical formula of accelerated and decelerated jet models \citep[see][]{Fan1994Ap&SS,Long2026MN}:

\begin{equation}
    \begin{cases}
    \delta_\nu=\delta_{\rm o}^{1+\frac{1}{8}\log \frac{\nu_{\rm o}}{\nu}} & (\alpha_{\rm rx}>0.75)\\
    \delta_{\rm 5GHz}=10^{\log(\frac{\delta_\nu}{\nu^{0.07}})+0.682} & (\alpha_{\rm rx}<0.75)
    \end{cases}\label{Eq-jetmode}
\end{equation}
In the first item, $\delta_{\rm o}$ is the optical Doppler factor at the typical optical band \citep[$\nu_{\rm o}=10^{14}$\,Hz; see][]{Fan1994Ap&SS}. This item is appropriate only to the accelerated jet model for strong-jet sources \citep[FSRQs+LBLs;][]{Fan1994Ap&SS,Liodakis2018ApJ}. The second item corresponds to the decelerated jet model of HBLs \citep{Long2026MN}. For IBLs, they may be the transitional objects between LBLs and HBLs \citep{Nieppola2006A&A,Abdo2010ApJ,Meyer2011ApJ}. Therefore, for IBLs with $\alpha_{\rm rx}>0.75$ ($\rm IBL_{>0.75}$), we apply the first item of Eq\,(\ref{Eq-jetmode}), otherwise (i.e., $\rm IBL_{<0.75}$), we apply the second item. Note that we consider Eq\,(\ref{Eq-jetmode}) to be applicable for extrapolations between radio bands. However, extrapolation across widely separated bands (e.g., $\delta_{\rm r}\rightarrow\delta_{\rm o}$ or $\delta_{\rm x}$) would introduce substantially larger uncertainties and was therefore avoided. If the sources have multiple $\delta_{\rm 5GHz}$, we will take the average values $\langle \delta_{\rm 5GHz}\rangle$ of them. Note that values of $\delta_{\rm r}<1$ are excluded, as they would imply the observational emissions ($f_\nu^{\rm ob}$) weaker than the intrinsic ones ($f_\nu^{\rm in}$; see Eq\,(\ref{Eq-beaming})), corresponding to Doppler de-boosting. Such a scenario is generally considered unlikely for blazars and is more likely to arise from the large uncertainties associated with estimates of $\delta_{\rm r}$. Excluding $\delta_{\rm r}<1$ measurements does not affect our results. This is because, in the initial cross-matched sample, only a very small fraction of the sources have reported $\delta_{\rm r}<1$ values. Among these sources, those with available broad-band data are still included in our sample; only the corresponding $\delta_{\rm r}<1$ values are excluded from the analysis.

$L_{\rm disk}$ is difficult to measure for blazars because their entire electromagnetic spectrum is dominated by non-thermal radiation. For blazars, $L_{\rm disk}$ is usually derived from the luminosity of broad line region (BLR), $L_{\rm BLR}$, by assuming $L_{\rm disk}=10L_{\rm BLR}$ \citep[e.g.,][]{Ghisellini2014Nat,Paliya2021ApJS}. However, most BL Lacs lack emission lines, which results in unavailable measurements of $L_{\rm disk}$. To obtain the more reliable $\lambda_{\rm Edd}$, we made our best effort to search for the $L_{\rm BLR}$ and $M_{\rm BH}$ of our blazars in the literatures (for the references see Table\,\ref{table1}). However, we find that there are 127/136 FSRQs, 34/64 LBLs, 8/39 IBLs, and 42/105 HBLs that have both available $L_{\rm disk}$ and $M_{\rm BH}$ in our sample (the composition of our sample is summarized in Table\,\ref{table3}). Excluding sources lacking both $L_{\rm disk}$ and $M_{\rm BH}$ may improve reader confidence, yet it would drastically reduce the BL Lac sample size, which would severely impact our statistical results (for more detailed discussions see Appendix\,\ref{appendix}). Therefore, for blazars without available $L_{\rm disk}$, we try to investigate a correlation between $L_{\rm disk}$ and other luminosities to estimate their $L_{\rm disk}$. Fortunately, we find a strong correlation between $L_{\rm disk}$ and $L_{\rm x}^{\rm ob}$ for FSRQs+LBLs+$\rm IBL_{>0.75}$ and HBLs+$\rm IBL_{<0.75}$, respectively (see Eq\,(\ref{Eq-A1}) and Eq\,(\ref{Eq-A2})). Therefore, for blazars without available $L_{\rm disk}$, we use Eq\,(\ref{Eq-A1}) and Eq\,(\ref{Eq-A2}) to estimate their $L_{\rm disk}$.

The $M_{\rm BH}$ of our blazars are taken from the existing literatures (see Table\,\ref{table1}). The traditional virial BH mass is also known as the dynamical mass, which adopts an empirical relationship between the size of BLR and the continuum luminosity, as well as the measured broad-line width, assuming that the BLR clouds gravitationally bound to the central BH \citep{Vestergaard2006ApJ}. This method is usually applied to estimate $M_{\rm BH}$ for FSRQs and some LBLs that show well-detected broad lines \citep[e.g.,][]{Shaw2012ApJ}. For HBLs, IBLs, and some LBLs, they have no or weak emission lines, so their $M_{\rm BH}$ are usually estimated from the properties of their host galaxies, e.g., $M_{\rm BH}-\sigma$ and $M_{\rm BH}-L_{\rm bulge}$ relations (where $\sigma$ and $L_{\rm bulge}$ are the stellar velocity dispersion and the bulge luminosity of the host galaxies, respectively). For blazars without available $M_{\rm BH}$, we used the fundamental plane (FP) of BH activity to estimate their $M_{\rm BH}$ \citep[e.g.,][]{Gultekin2009ApJ,Bariuan2022MN}. The Appendix\,\ref{appendix} provides the relevant knowledge about FP. Similarly, we divide our blazars into FSRQs+LBLs+$\rm IBL_{>0.75}$ and HBLs+$\rm IBL_{<0.75}$; the best fits of their FPs are Eq\,(\ref{Eq-A4}) and Eq\,(\ref{Eq-A5}), respectively. Therefore, for blazars without available $M_{\rm BH}$, we use Eq\,(\ref{Eq-A4}) and Eq\,(\ref{Eq-A5}) to estimate their $M_{\rm BH}$. Note that the means above do not have a significant effect on our results in the next (see Appendix\,\ref{appendix}).

\begin{figure*}
    \centering
    \includegraphics[width=\textwidth]{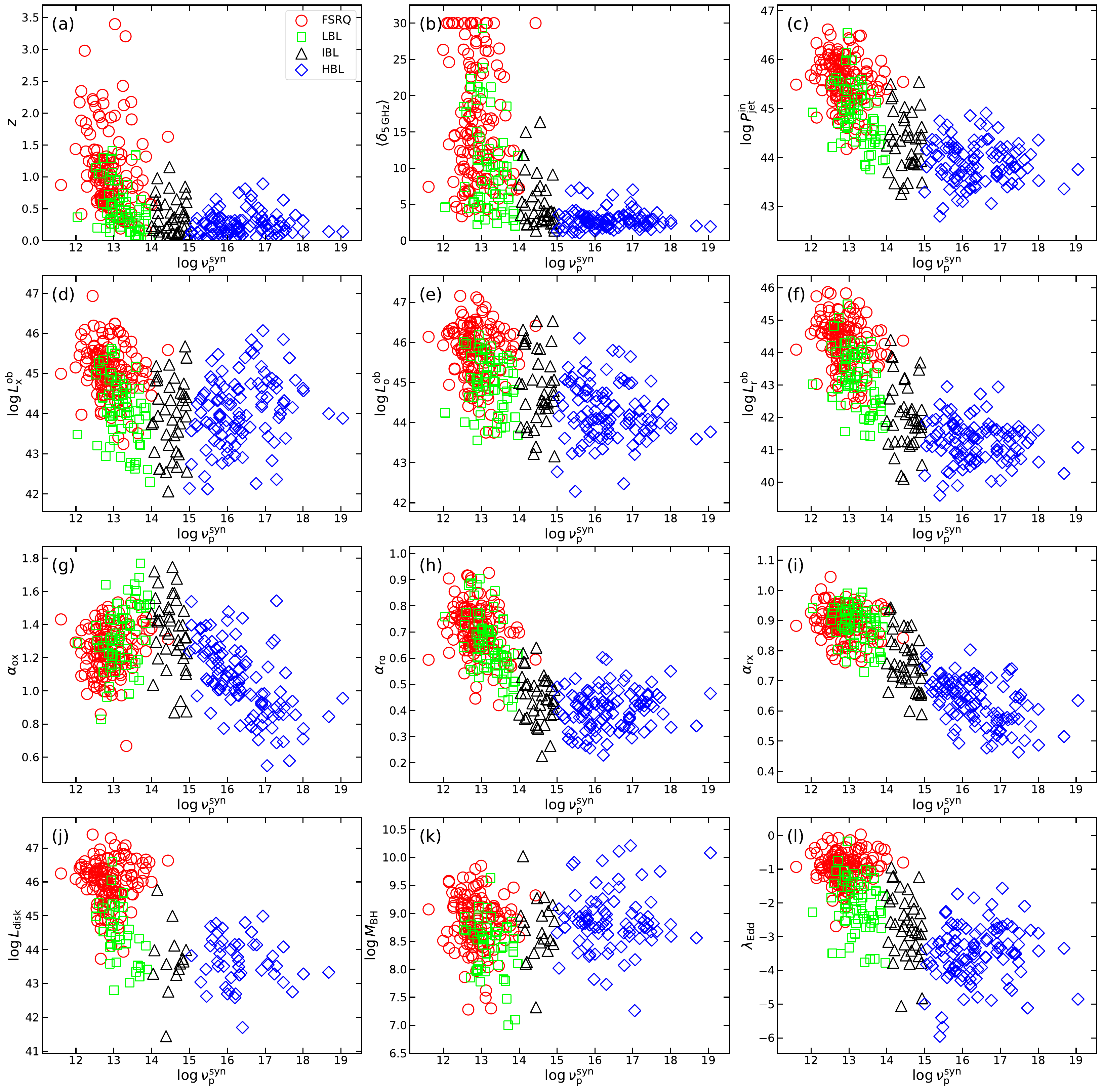}
    \caption{The relation between $\nu_{\rm p}^{\rm syn}$ and properties for the blazar sequence. The red circles, lime squares, black triangles, and blue diamonds are FSRQs, LBLs, IBLs, and HBLs, respectively. (a) $z$ vs. $\log \nu_{\rm p}^{\rm syn}$ diagram. (b) $\langle \delta_{\rm 5GHz}\rangle$ vs. $\log \nu_{\rm p}^{\rm syn}$ diagram. (c) $\log P_{\rm jet}^{\rm in}$ vs. $\log \nu_{\rm p}^{\rm syn}$ diagram. (d) $\log L_{\rm x}^{\rm ob}$ vs. $\log \nu_{\rm p}^{\rm syn}$ diagram. (e) $\log L_{\rm o}^{\rm ob}$ vs. $\log \nu_{\rm p}^{\rm syn}$ diagram. (f) $\log L_{\rm r}^{\rm ob}$ vs. $\log \nu_{\rm p}^{\rm syn}$ diagram. (g) $\alpha_{\rm ox}$ vs. $\log \nu_{\rm p}^{\rm syn}$ diagram. (h) $\alpha_{\rm ro}$ vs. $\log \nu_{\rm p}^{\rm syn}$ diagram. (i) $\alpha_{\rm rx}$ vs. $\log \nu_{\rm p}^{\rm syn}$ diagram. (j) $\log L_{\rm disk}$ vs. $\log \nu_{\rm p}^{\rm syn}$ diagram. (k) $\log M_{\rm BH}$ vs. $\log \nu_{\rm p}^{\rm syn}$ diagram. (l) $\lambda_{\rm Edd}$ vs. $\log \nu_{\rm p}^{\rm syn}$ diagram.}
    \label{fig-(V-p)}
\end{figure*}

The basic parameters of first sources of FSRQs, LBLs, IBLs, and HBLs are given in the representative Table\,\ref{table1}. The complete table is available online in machine-readable format. Figure\,\ref{fig-(V-p)} shows the distributions and comparisons of properties for our blazars.

\begin{longrotatetable}
\begin{deluxetable*}{ccccccccccccccccc}
		\tablecaption{The Properties of Blazars\label{table1}}
		\renewcommand{\tabcolsep}{1.85pt}
		\tablehead{
			\colhead{IAU Name} & \colhead{$z$} & 
			\colhead{$\log \nu_{\rm p}^{\rm syn}$} & \colhead{Class} & 
			\colhead{$\log L_{\rm x}^{\rm ob}$} & \colhead{$\log L_{\rm o}^{\rm ob}$} &  \colhead{$\log L_{\rm r}^{\rm ob}$} & 
			\colhead{$\log P_{\rm jet}^{\rm in}$} & \colhead{$\log L_{\rm disk}$} & \colhead{$\log M_{\rm BH}$} & \colhead{Refs.} & \colhead{$\alpha_{\rm ox}$} & \colhead{$\alpha_{\rm ro}$} & \colhead{$\alpha_{\rm rx}$} & \colhead{$\delta_{\rm 5GHz}$} & \colhead{$\langle \delta_{\rm 5GHz}\rangle$} & \colhead{$\lambda_{\rm Edd}$} \\ 
			\colhead{(1)} & \colhead{(2)} & \colhead{(3)} & \colhead{(4)} & 
			\colhead{(5)} & \colhead{(6)} &
			\colhead{(7)} & \colhead{(8)} & \colhead{(9)} & \colhead{(10)} & \colhead{(11)} & \colhead{(12)} & \colhead{(13)} & \colhead{(14)} & \colhead{(15)} & \colhead{(16)} & \colhead{(17)}
		} 
		\startdata
0007$+$205 & 0.600 & 12.56 & FSRQ & 44.48 & 44.70 & 42.83 & 44.90 & 45.35 & 7.86 & 1/2/1/2/3/3 & 1.081 & 0.630 & 0.785 & .../6.47/.../.../... & 6.47 & $-0.624$ \\
... & ... & ... & ... & ... & ... & ... & ... & ... & ... & .../.../.../.../.../... & ... & ... & ... & .../.../.../.../... & ... & ... \\
0003$-$066 & 0.347 & 12.92 & LBL & 44.05 & 44.91 & 43.58 & 45.16 & 44.52 & 8.93 & 1/2/1/2/3/3 & 1.323 & 0.737 & 0.939 & 5.40/7.53/.../5.97/5.58 & 6.12 & $-2.524$ \\
... & ... & ... & ... & ... & ... & ... & ... & ... & ... & .../.../.../.../.../... & ... & ... & ... & .../.../.../.../... & ... & ... \\
0048$-$09 & 0.634 & 14.12 & IBL & 45.18 & 45.97 & 43.89 & 45.21 & ... & 8.85 & 1/2/9/2/.../3 & 1.296 & 0.587 & 0.832 & 8.46/22.84/12.10/4.58/10.88 & 11.77 & $-0.948$ \\
... & ... & ... & ... & ... & ... & ... & ... & ... & ... & .../.../.../.../.../... & ... & ... & ... & .../.../.../.../... & ... & ... \\
0032$+$595 & 0.086 & 17.05 & HBL & 44.40 & 43.21 & 40.61 & 43.75 & ... & 7.26 & 1/2/9/2/.../29 & 0.549 & 0.484 & 0.506 & 3.19/.../.../2.33/... & 2.76 & $-1.565$ \\
... & ... & ... & ... & ... & ... & ... & ... & ... & ... & .../.../.../.../.../... & ... & ... & ... & .../.../.../.../... & ... & ... \\
         \enddata
            
		\tablecomments{Col\,(1): IAU Name. Col\,(2): Redshift. Col\,(3): logarithm of $\nu_{\rm p}^{\rm syn}$ for blazars (Hz). Col\,(4): the classes of blazars. Col\,(5): logarithm of the observational 1\,keV X-ray luminosity ($L_{\rm x}^{\rm ob}$ in units of $\rm erg\,s^{-1}$). Col\,(6): logarithm of the observational optical luminosity at $5.4143\times10^{14}$\,Hz $V$-band ($L_{\rm o}^{\rm ob}$ in units of $\rm erg\,s^{-1}$). Col\,(7): logarithm of the observational 5\,GHz core radio luminosity ($L_{\rm x}^{\rm ob}$ in units of $\rm erg\,s^{-1}$). Col\,(8): logarithm of the intrinsic jet power ($P_{\rm jet}^{\rm in}$ in units of $\rm erg\,s^{-1}$). Col\,(9): logarithm of the accretion disk luminosity ($L_{\rm disk}$ in units of $\rm erg\,s^{-1}$). Col\,(10): logarithm of the dynamical BH mass ($M_{\rm BH}$ in units of $M_{\odot}$). Col\,(11): references of the X-ray flux, optical flux, 5 GHz core radio flux, low-frequency radio flux, accretion disk luminosity, and BH mass. Col\,(12), Col\,(13), and Col\,(14) are the broad-band spectral indices from optical $5.4143\times10^{14}$\,Hz $V$-band to 1\,keV X-ray band ($\alpha_{\rm ox}$), from 5\,GHz radio band to optical $5.4143\times10^{14}$\,Hz $V$-band ($\alpha_{\rm ro}$), and from 5\,GHz radio band to 1\,keV X-ray band ($\alpha_{\rm rx}$), respectively. Col\,(15): the 5\,GHz radio Doppler factors refer to \cite{Ye2021PASJ}/\cite{Liodakis2018ApJ}/\cite{Liodakis2017MN}/\cite{Wu2014}/\cite{Hovatta2009} (Note, $\delta_{\rm r}$ at other radio band have been extrapolated to $\delta_{\rm 5GHz}$). Col\,(16): the average 5\,GHz Doppler factor $\langle \delta_{\rm 5GHz}\rangle$. Col\,(17): the Eddington ratio $\lambda_{\rm Edd}$.\\
    References: (1)\,NED; (2)\,CDS; (3)\,\cite{Paliya2021ApJS}; (4)\,\cite{Chen2024ApJS}; (5)\,\cite{Sikora2007ApJ}; (6)\,\cite{Chai2012ApJ}; (7)\,\cite{Zhou2009RAA}; (8)\,\cite{Saxton2008}; (9)\,\cite{Ye2021PASJ}; (10)\,\cite{Liu2006ApJ}; (11)\,\cite{Xiong2015MN-a}; (12)\,\cite{Wang2006ApJ}; (13)\,\cite{Dwelly2017MN}; (14)\,\cite{Shen2011ApJS}; (15)\,\cite{Kurinsky2013}; (16)\,\cite{Yuan2012ApJ}; (17)\,\cite{Gu2009MN}; (18)\,\cite{Mantovani2015}; (19)\,\cite{Long2025ApJ}; (20)\,\cite{Marshall2018ApJ}; (21)\,\cite{Sbarrato2012MN}; (22)\,\cite{Yuan2018ApJS}; (23)\,\cite{Kovalev2020MN}; (24)\,\cite{Woo2002ApJ}; (25)\,\cite{Pian2005MN}; (26)\,$M_{\rm R}$ from \cite{Nilsson2003}, with $M_{\rm BH}$ estimated via the $M_{\rm BH}-L_{\rm bulge}$ relation of \cite{Graham2007MN}; (27)\,\cite{Brinkmann1997}; (28)\,\cite{Ghisellini2011MN}; (29)\,\cite{Wu2009RAA}; (30)\,\cite{Wu2014}; (31)\,\cite{Fan2004ApJ}; (32)\,$M_{\rm R}$ from \cite{O-Dowd2005ApJ}, with $M_{\rm BH}$ estimated via the $M_{\rm BH}-L_{\rm bulge}$ relation of \cite{Graham2007MN}; (33)\,\cite{Plotkin2011MN}; (34)\,\cite{Falomo2003ApJ}; (35)\,\cite{Long2026MN}; (36)\,\cite{Wang2002ApJ}.}
	\end{deluxetable*}
\end{longrotatetable}

\begin{deluxetable*}{cccccccccccc}
\tablecaption{The Average Properties for Blazars\label{table2}}
		\renewcommand{\tabcolsep}{6pt}
    \tablehead{Class & $\langle \log \nu_{\rm p}^{\rm syn}\rangle$ & $\langle z\rangle$ & $\langle \alpha_{\rm ox}\rangle$ & $\langle \alpha_{\rm ro}\rangle$ & $\langle \alpha_{\rm rx}\rangle$ & $\langle \log L_{\rm x}^{\rm ob}\rangle$ & $\langle \log L_{\rm o}^{\rm ob}\rangle$ & $\langle \log L_{\rm r}^{\rm ob}\rangle$ & $\langle \log P_{\rm jet}^{\rm in}\rangle$ & $\langle \delta_{\rm 5GHz}\rangle$ & $\langle \lambda_{\rm Edd}\rangle$ }

	\startdata
	FSRQ & 12.89 & 1.094 & 1.248 & 0.709 & 0.895 & 45.07 & 45.73 & 44.27 & 45.57 & 15.90 & $-0.959$ \\
    LBL & 13.21 & 0.515 & 1.338 & 0.646 & 0.884 & 44.05 & 44.95 & 43.16 & 44.86 & 10.09 & $-2.089$ \\
	IBL & 14.51 & 0.380 & 1.356 & 0.445 & 0.759 & 43.92 & 44.87 & 42.07 & 44.33 & 5.31 & $-2.695$ \\
    HBL & 16.38 & 0.250 & 1.060 & 0.404 & 0.630 & 44.09 & 44.24 & 41.24 & 43.89 & 2.84 & $-3.490$ \\
    \enddata   
\end{deluxetable*}

\begin{figure*}
\centering  
\includegraphics[width=0.65\textwidth]{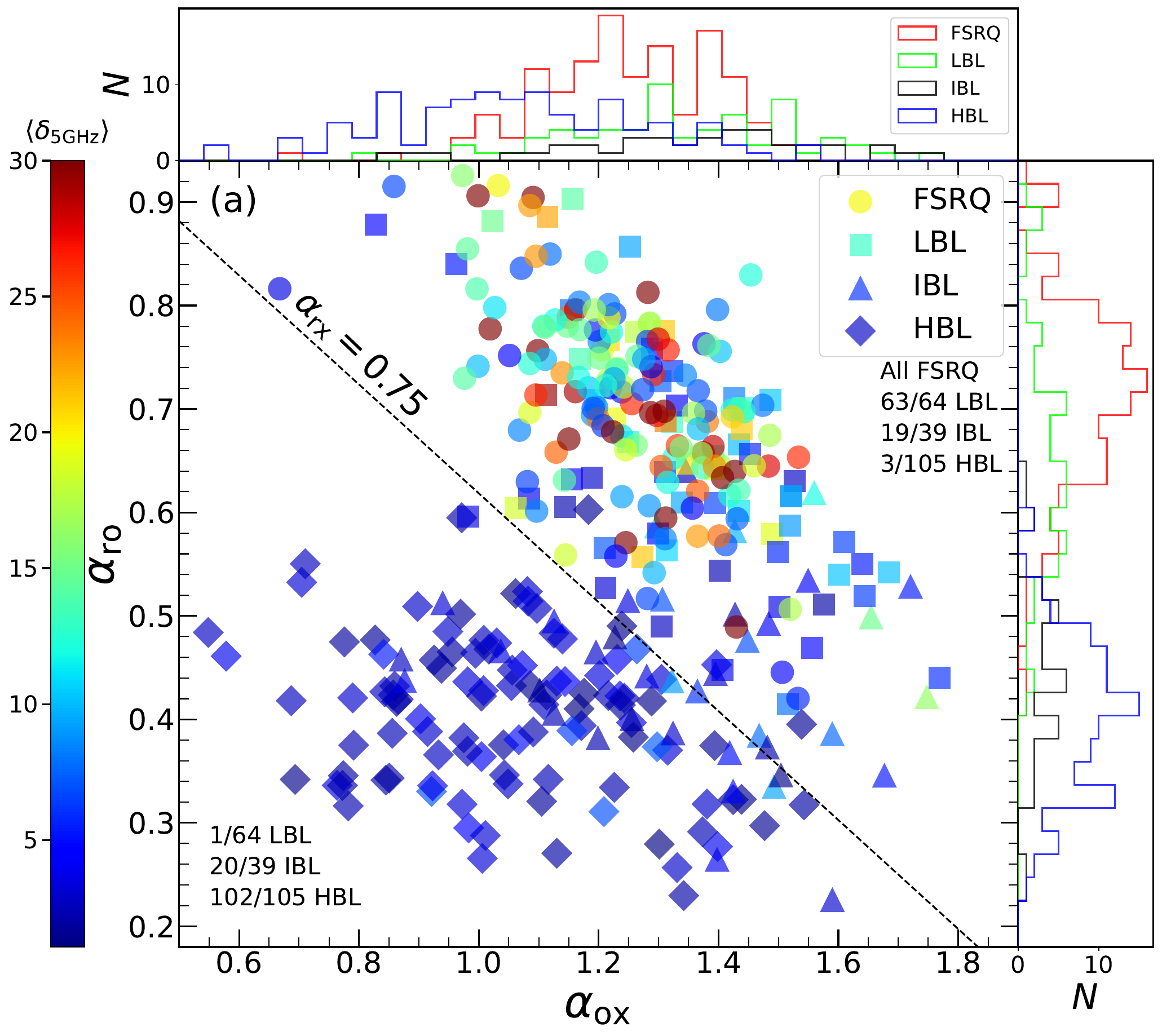}\\
\includegraphics[width=1\textwidth]{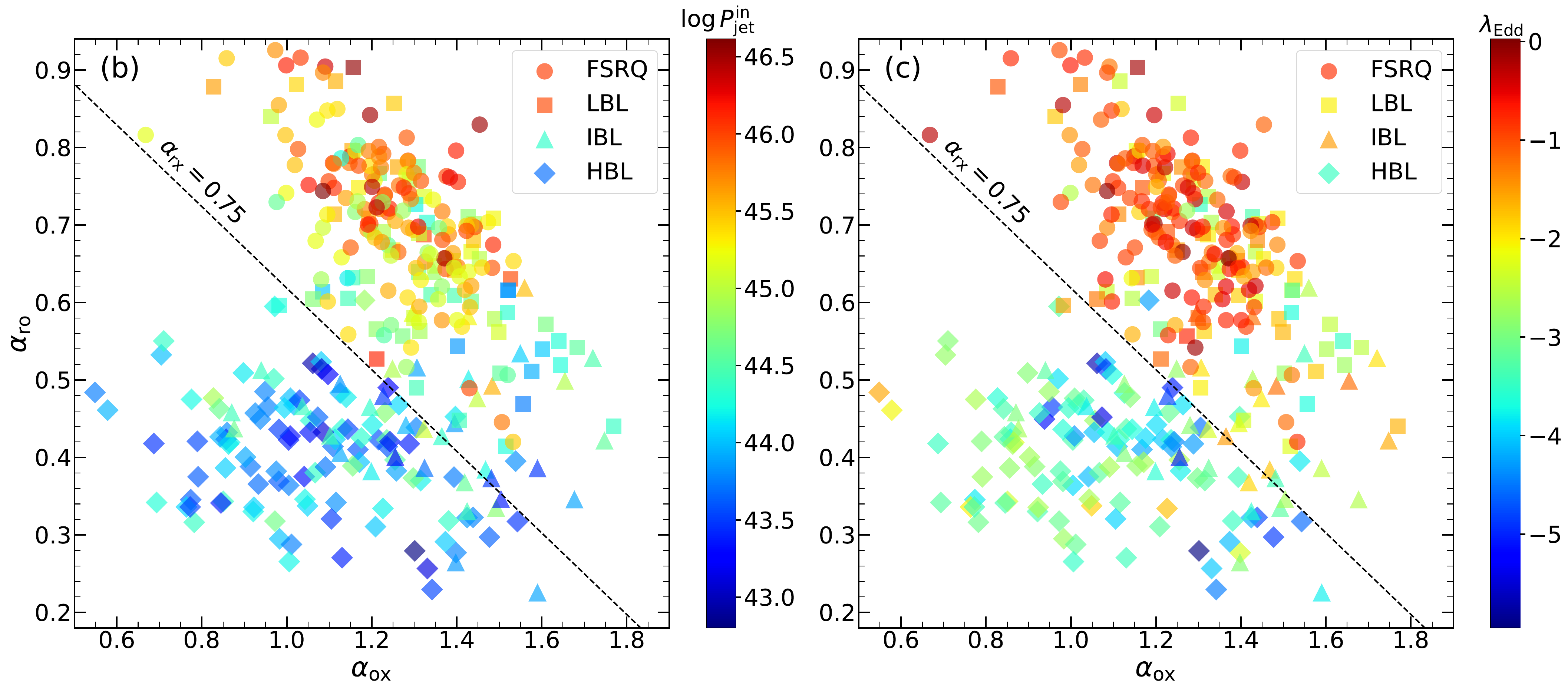}
\caption{The $\alpha_{\rm ro}$ vs. $\alpha_{\rm ox}$ diagram. The circles, squares, triangles, and diamonds are FSRQs, LBLs, IBLs, and HBLs, respectively. $\alpha_{\rm rx}$ can be regarded as the intercept; the sources that occupy the upper right region have $\alpha_{\rm rx}>0.75$, on the contrary, the sources that occupy the lower left corner have $\alpha_{\rm rx}<0.75$. (a), (b), and (c) show Doppler beaming effect at 5 GHz $\langle \delta_{\rm 5GHz}\rangle$, intrinsic jet power $\log P_{\rm jet}^{\rm in}$, and accretion rates $\lambda_{\rm Edd}$, respectively, with color scales representing each quantity. The histograms on the axes in plane-(a) show the distribution of these physical quantities ($\alpha_{\rm ox}$ and $\alpha_{\rm ro}$). The red, lime, black, and blue boxes are FSRQs, LBLs, IBLs, and HBLs, respectively.}
\label{fig-index}
\end{figure*}

\begin{figure*}
    \centering{\includegraphics[scale=0.399]{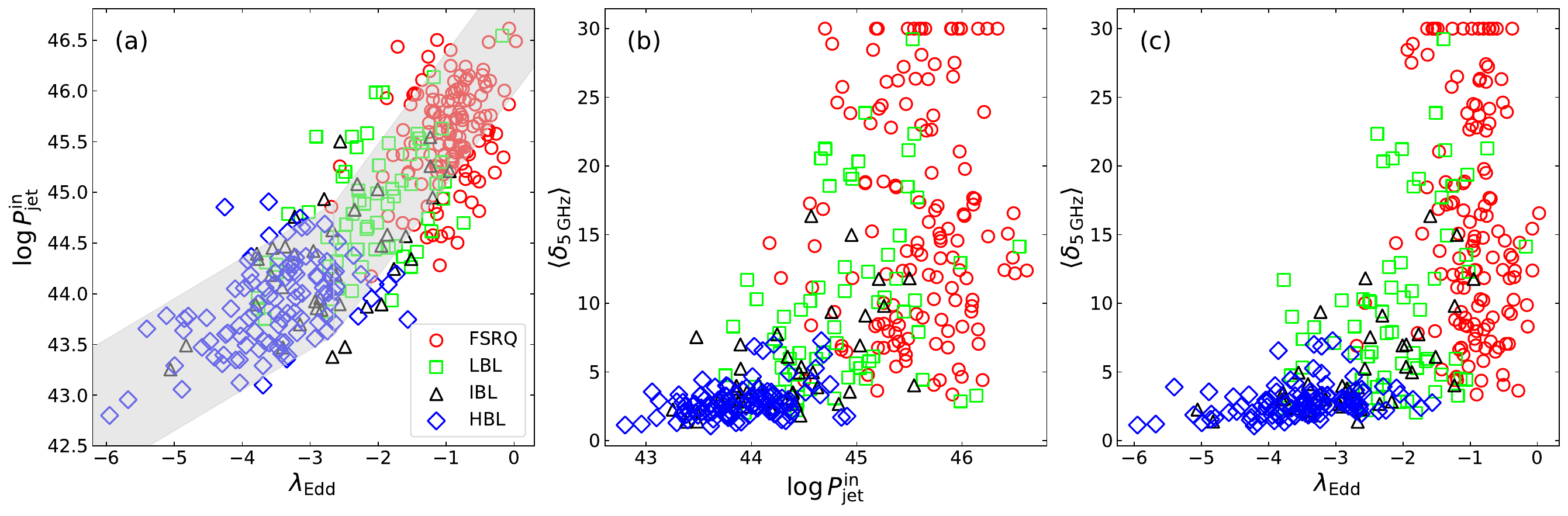}}
    \caption{The comparison between intrinsic jet power ($\log P_{\rm jet}^{\rm in}$), Eddington radio ($\lambda_{\rm Edd}$), and 5 GHz Doppler factor ($\langle \delta_{\rm 5GHz}\rangle$) for our sample. (a) $\log P_{\rm jet}^{\rm in}$ vs. $\lambda_{\rm Edd}$ diagram; (b) $\langle \delta_{\rm 5GHz}\rangle$ vs. $\log P_{\rm jet}^{\rm in}$ diagram; (c) $\langle \delta_{\rm 5GHz}\rangle$ vs. $\lambda_{\rm Edd}$ diagram. The notations are the same as that of Figure\,\ref{fig-(V-p)}.}
    \label{fig-(P+Edd+D)}
\end{figure*}

\section{Results}\label{Sec-result}

\subsection{The Distribution and Comparison of Properties among Blazars}

The underlying physics of observation properties for blazars can be obtained from the fundamental physics parameters, which can reveal the similarities and differences among them. Table\,\ref{table2} provides the average properties of blazars. Figure\,\ref{fig-(V-p)} shows the distributions and comparisons of parameters for our blazars.

Table\,\ref{table2} shows that all average properties, except $\alpha_{\rm ox}$ and $\log L_{\rm x}^{\rm ob}$, are negatively correlated with $\nu_{\rm p}^{\rm syn}$ among blazar subclasses.

Figure\,\ref{fig-(V-p)}-(a) shows a rough anti-correlation between redshift and $\log \nu_{\rm p}^{\rm syn}$ for blazars, with a Pearson correlation coefficient of $R=-0.547$ and $P=3.01\times10^{-28}$. This finding is consistent with \cite{Fossati1998MN}, suggesting that FSRQs are more readily detected at higher $z$.

Figure\,\ref{fig-(V-p)}-(b) shows a rough anti-correlation between $\langle \delta_{\rm 5GHz}\rangle$ and $\log \nu_{\rm p}^{\rm syn}$ for the blazar sequence, with a Pearson correlation coefficient of $R=-0.621$ and $P=4.65\times10^{-38}$. This finding is consistent with \cite{Wu2007A&A}, \cite{Nieppola2008A&A}, \cite{Yang2022ApJ}, and \cite{Long2025ApJ}. Combining the VLBA observations discussed in Section\,\ref{Sec-intro}, the observed $\delta_{\rm r}-\nu_{\rm p}^{\rm syn}$ anti-correlation can be interpreted in terms of an accelerating/decelerating jet scenario across different blazar subclasses. At least for HBLs, their jets may have already decelerated to subrelativistic speeds and become decollimated by the time they propagate to the radio region, whereas the jets in FSRQs and most LBLs still maintain highly relativistic speeds on radio scales \citep[$\delta \approx \Gamma$; see][]{Ghisellini2014Nat}. Note there are 15 FSRQs that have $\langle \delta_{\rm 5GHz}\rangle>30$, with a maximum value of $\langle \delta_{\rm 5GHz}\rangle=105.97$ for 0446$+$112. To mitigate the significant disparity and obtain a more continuous distribution of $\langle \delta_{\rm 5GHz}\rangle$ in Figure\,\ref{fig-(V-p)}-(b), \ref{fig-index}, and \ref{fig-(P+Edd+D)}, we assigned a cap value of $\langle \delta_{\rm 5GHz}\rangle=30$ to these extreme outliers. This measure does not have a significant effect on our results.

Figure\,\ref{fig-(V-p)}-(c) shows an anti-correlation between $\log P_{\rm jet}^{\rm in}$ and $\log \nu_{\rm p}^{\rm syn}$ for blazar sequence, with a Pearson correlation coefficient of $R=-0.757$ and $P=3.31\times10^{-65}$. This suggests different jet strengths among blazars, with blazars having lower $\nu_{\rm p}^{\rm syn}$ exhibiting more powerful jets.

Figure\,\ref{fig-(V-p)}-(d) shows a $\lor$-type evolution in the $\log L_{\rm x}^{\rm ob}-\log \nu_{\rm p}^{\rm syn}$ relation for the blazar sequence, which agrees with \cite{Nieppola2006A&A}, \cite{Fan2016ApJS}, and \cite{Yang2022ApJS}. The break occurs at about $\nu_{\rm p}^{\rm syn}\sim10^{14.5}$\,Hz, where IBLs seem to serve as a bridge between FSRQs+LBLs and HBLs.

Figure\,\ref{fig-(V-p)}-(e) shows an anti-correlation between $\log L_{\rm o}^{\rm ob}$ and $\log \nu_{\rm p}^{\rm syn}$ for the blazar sequence, with a Pearson correlation coefficient of $R=-0.603$ and $P=2.09\times10^{-35}$.

Figure\,\ref{fig-(V-p)}-(f) shows an anti-correlation between $\log L_{\rm r}^{\rm ob}$ and $\log \nu_{\rm p}^{\rm syn}$ for the blazar sequence, with a Pearson correlation coefficient of $R=-0.803$ and $P=6.38\times10^{-79}$, which is consistent with \cite{Fossati1998MN}, \cite{Nieppola2006A&A}, \cite{Fan2016ApJS}, and \cite{Yang2022ApJS}.

Figure\,\ref{fig-(V-p)}-(g) shows a $\land$-type evolution in the $\alpha_{\rm ox}-\log \nu_{\rm p}^{\rm syn}$ relation for the blazar sequence (here, IBLs also seem to serve as a bridge between FSRQs+LBLs and HBLs), which is opposite to the $\log L_{\rm x}^{\rm ob}-\log \nu_{\rm p}^{\rm syn}$ relation. This finding is consistent with \cite{Fan2016ApJS} and \cite{Yang2022ApJS}.

Figure\,\ref{fig-(V-p)}-(h) shows an anti-correlation between $\alpha_{\rm ro}$ and $\log \nu_{\rm p}^{\rm syn}$ in the sequence of FSRQs→LBLs→IBLs, with a Pearson correlation coefficient of $R=-0.714$ and $P=1.39\times10^{-38}$. HBLs show an almost horizontal line in the $\alpha_{\rm ro}-\log \nu_{\rm p}^{\rm syn}$ relation. This finding roughly agrees with \cite{Fossati1998MN}, \cite{Fan2016ApJS}, and \cite{Yang2022ApJS}.

Figure\,\ref{fig-(V-p)}-(i) shows an anti-correlation between $\alpha_{\rm rx}$ and $\log \nu_{\rm p}^{\rm syn}$ for the blazar sequence, with a Pearson correlation coefficient of $R=-0.875$ and $P=7.26\times10^{-110}$. This finding is consistent with \cite{Fossati1998MN}.

Figure\,\ref{fig-(V-p)}-(j) shows an anti-correlation between $\log L_{\rm disk}$ and $\log \nu_{\rm p}^{\rm syn}$ for blazars with available $L_{\rm disk}$, with a Pearson correlation coefficient of $R=-0.438$ and $P=9.63\times10^{-23}$.

Figure\,\ref{fig-(V-p)}-(k) shows an almost horizontal line in the $\log M_{\rm BH}-\log \nu_{\rm p}^{\rm syn}$ relation for blazars with available $M_{\rm BH}$, with a Pearson correlation coefficient of $R=0.094$ and $P=0.115$.

Figure\,\ref{fig-(V-p)}-(l) shows an anti-correlation between $\lambda_{\rm Edd}$ and $\log \nu_{\rm p}^{\rm syn}$ for the blazar sequence, with a Pearson correlation coefficient of $R=-0.736$ and $P=5.06\times10^{-60}$. This finding agrees with \cite{Paliya2021ApJS}, suggesting different accretion regimes among blazars. The blazars with a lower $\nu_{\rm p}^{\rm syn}$ have higher accretion rates.

\subsection{The Distribution of the Doppler Beaming Effect, Jet Power, and Accretion Rates on the $\alpha_{\rm ro}-\alpha_{\rm ox}$ Plane}

To achieve our aims (see section\,\ref{Sec-intro}), we put our blazars into the $\alpha_{\rm ro}-\alpha_{\rm ox}$ plane and confer them with $\langle \delta_{\rm 5GHz}\rangle$, $P_{\rm jet}^{\rm in}$, and $\lambda_{\rm Edd}$ values in the scale of color; our results are shown in Figure\,\ref{fig-index}.

It is clear that FSRQs+LBLs$\rightarrow$IBLs$\rightarrow$HBLs follow a \reflectbox{$\angle$}-shaped evolutionary trajectory, which is consistent with previous studies \citep[e.g.,][]{Padovani1995ApJ,Donato2001A&A,Abdo2010ApJ,Fan2016ApJS,Yang2022ApJS}. In the $\alpha_{\rm ro}$--$\alpha_{\rm ox}$ plane, $\alpha_{\rm rx}$ can be regarded as the intercept \citep[see][]{Ledden1985ApJ}. In other words, $\alpha_{\rm rx}$ increases toward the upper-right of the plane. Our results show that all FSRQs reside in the URR with $\alpha_{\rm rx}>0.75$, there are 63 out of 64 (63/64) LBLs that occupy the URR, only 1/64 LBL occupies the lower left region (LLR) with $\alpha_{\rm rx}<0.75$. In addition, the region inhabited by FSRQs significantly overlaps with that of LBLs. IBLs are almost equally distributed between URR (19/39) and LLR (20/39). There are 102/105 HBLs that occupy LLR, only 3/105 HBLs occupy URR. IBLs seem to be the transitional objects between FSRQs+LBLs and HBLs. The statistics above are presented in Figure\,\ref{fig-index}-(a).

The histograms on the axes of Figure\,\ref{fig-index}-(a) show the distribution of $\alpha_{\rm ox}$ and $\alpha_{\rm ro}$. On average, the distribution of $\alpha_{\rm ox}$ of FSRQs+LBLs is only slightly larger than that of HBLs. However, the distributions of $\alpha_{\rm ro}$ and $\alpha_{\rm rx}$ of FSRQs+LBLs are significantly larger than those of HBLs (see also Figure\,\ref{fig-(V-p)} (g), (h), and (i)).

From Figure\,\ref{fig-index}-(a), we find a nearly continuous distribution of $\langle \delta_{\rm 5GHz}\rangle$ for the blazar sequence, in which $\langle \delta_{\rm 5GHz}\rangle$ decreases toward the \reflectbox{$\angle$}-shaped track. $P_{\rm jet}^{\rm in}$ and $\lambda_{\rm Edd}$ also decrease toward this track (see Figures\,\ref{fig-index} (b) and (c)). It is clear that there are the positive correlations among $\langle \delta_{\rm 5GHz}\rangle$, $P_{\rm jet}^{\rm in}$ and $\lambda_{\rm Edd}$. For a more intuitive illustration, we present the plots in Figure\,\ref{fig-(P+Edd+D)}. Figure\,\ref{fig-(P+Edd+D)}-(a) shows a break at $\lambda_{\rm Edd}\sim-2.651$ in the $\log P_{\rm jet}^{\rm in}-\lambda_{\rm Edd}$ relation, which is estimated by visual inspection. This critical value is consistent with the theoretical transition value between the radiatively efficient and radiatively inefficient accretion regimes \citep[e.g.,][]{Ghisellini2008MN,Yuan2014ARA&A}.

\section{Discussion}\label{Sec-discu}

\begin{figure*}
    \centering{\includegraphics[scale=0.61]{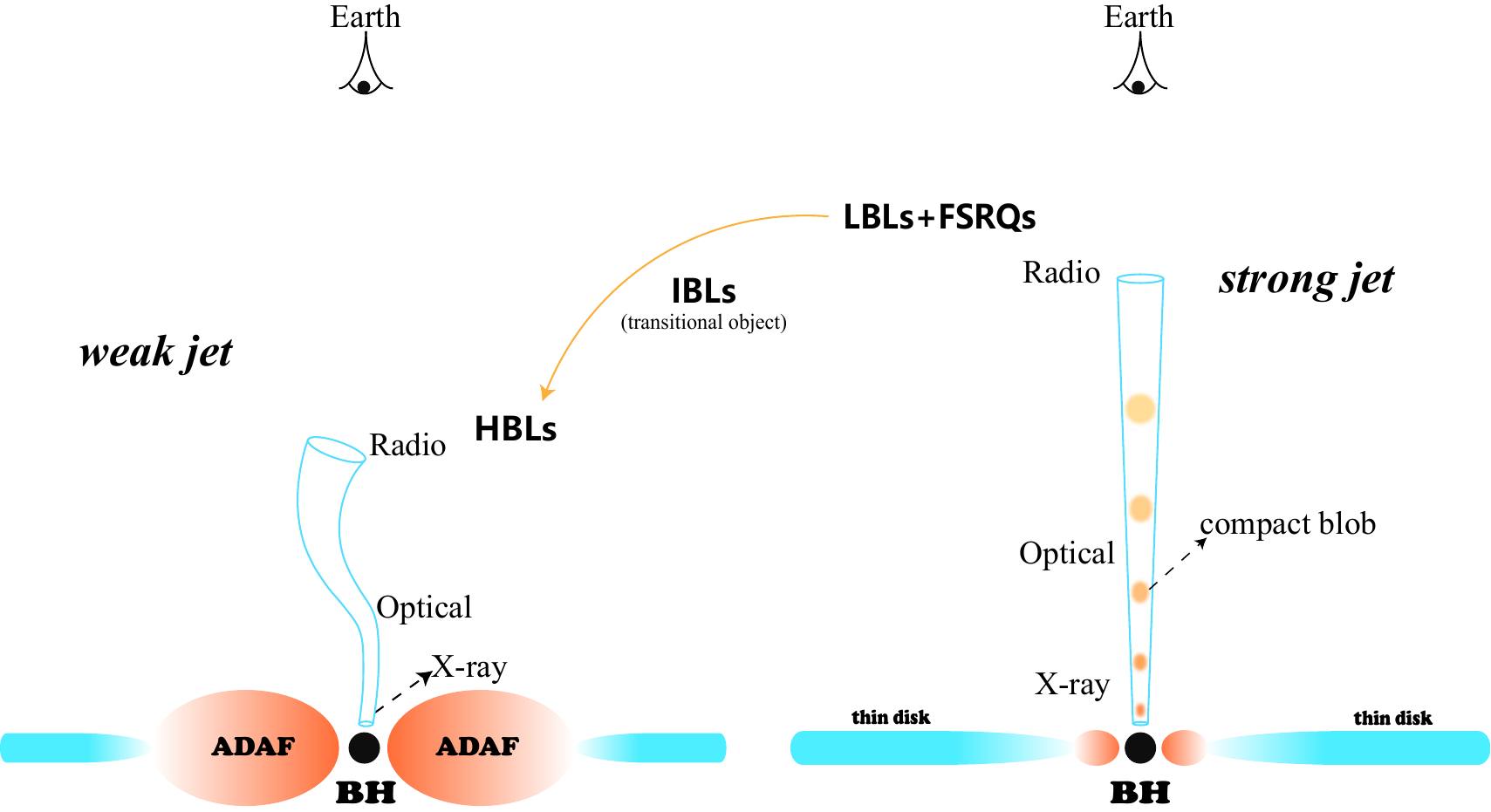}}
    \caption{The cartoon picture of blazar sequence. This accretion disk mode refers from the gradual evolution of accretion disk in BH-XRB \citep[e.g.,][]{Fender2004MN,Ferreira2006,Fender2012Sci,Begelman2014ApJL}. Noting, the accretion disk in the left is dominated by ADAF instead of the hybrid mode of ADAF+SSD. The main difference of jets between FSRQs+LBLs and HBLs occur in the region from optical to radio. In addition, the jet bases of FSRQs+LBLs are capable of producing compact blobs, which move ultra-relativistically along the jet axis.}
    \label{fig-sequence}
\end{figure*}

Before proceeding with further discussion, we make the following assumptions, which have also been adopted in many previous works \citep[e.g.,][]{Chiaberge2000A&A,Raiteri2017Natur}. We assume that the radio, optical, and X-ray emissions are dominated by jets, and they come from different regions along the jet, as shown in Figure\,\ref{fig-sequence} \citep[see also][]{Raiteri2017Natur}.

\subsection{The Physical Properties behind the Distribution of Blazars on the $\alpha_{\rm ro}-\alpha_{\rm ox}$ plane}\label{Sec-4.1}

Figure\,\ref{fig-index} implies that the distribution of blazars on the $\alpha_{\rm ro}-\alpha_{\rm ox}$ plane is regulated by the accretion rates and the Doppler beaming effect. Next, we will conduct a further analysis from a theoretical perspective. In the beaming model, $f_\nu^{\rm ob}$ is strongly boosted from $f_\nu^{\rm in}$, an empirical equation is given as:

\begin{equation}
    f_\nu^{\rm ob}=\delta_\nu^q f_\nu^{\rm in}
    \label{Eq-beaming}
\end{equation}
where $q=2+\alpha_\nu$ for the continuous jet and $q=3+\alpha_\nu$ for the spherical jet \citep{Scheuer1979Nat}, and $\alpha_\nu$ is the spectral index at $\nu$ band. $\delta_\nu$ is the Doppler factor at $\nu$ band. Combining Eq\,(\ref{Eq-alpha}) with Eq\,(\ref{Eq-beaming}), we have:

\begin{equation}
    \alpha_{ij}=-\log[(f_i^{\rm in} \delta_i^{q_i})/(f_j^{\rm in} \delta_j^{q_j})]/\log(\nu_i/\nu_j)
    \label{Eq-index(in)}
\end{equation}
Taking $\nu_{\rm r}=5\times10^9$\,Hz, $\nu_{\rm o}=5.4143\times10^{14}$\,Hz, and $\nu_{\rm x}=2.41799\times10^{17}$\,Hz into Eq\,(\ref{Eq-index(in)}), we can obtain:

\begin{equation}
    \begin{cases}
    \alpha_{\rm ro}=0.199\log(\frac{f_{\rm r}^{\rm in}\delta_{\rm r}^{q_{\rm r}}}{f_{\rm o}^{\rm in}\delta_{\rm o}^{q_{\rm o}}})\\
    \alpha_{\rm ox}=0.377\log(\frac{f_{\rm o}^{\rm in}\delta_{\rm o}^{q_{\rm o}}}{f_{\rm x}^{\rm in}\delta_{\rm x}^{q_{\rm x}}})\\
    \alpha_{\rm rx}=0.130\log(\frac{f_{\rm r}^{\rm in}\delta_{\rm r}^{q_{\rm r}}}{f_{\rm x}^{\rm in}\delta_{\rm x}^{q_{\rm x}}})
    \end{cases}\label{Eq-index(ratio)}
\end{equation}
It is clear that $\alpha_{ij}$ is determined by the ratio of $f_i^{\rm in}/f_j^{\rm in}$ and $\delta_i^{q_i}/\delta_j^{q_j}$, where $\delta_i^{q_i}/\delta_j^{q_j}$ is determined by $\delta_i/\delta_j$, the jet shape ($q=2(3)+\alpha_\nu$), and $\alpha_\nu$. 

In the theoretical model, the relativistic jets are produced from the innermost regions of the accretion disk, so the jet variables should depend on two fundamental parameters that determine the conditions in the inner accretion disk, namely the dimensionless accretion rate ($\dot{m}=\dot{M}/\dot{M}_{\rm Edd}$, where $\dot{M}_{\rm Edd}$ is the Eddington accretion rate) and $M_{\rm BH}$ \citep{Heinz2003MN}:

\begin{equation}
    f_\nu^{\rm in}\propto M_{\rm BH}^{\xi_M} \dot{m}^{\xi_{\dot{m}}} \propto \dot{M}^{\xi_\nu}
\end{equation}
this gives rise to the scaling relation:

\begin{equation}
    f_i^{\rm in}/f_j^{\rm in} \propto \dot{M}^{\xi_{ij}}
\end{equation}
where $\xi_{ij}$ is regulated by the accretion mode \citep[see][]{Heinz2003MN,Merloni2003MN,Yuan2005ApJ,Long2025ApJ}. Therefore, our observational results are consistent with the theoretical expectation.

However, $f_i^{\rm in}/f_j^{\rm in}$ of blazars are very difficult to obtain due to the fact that $\delta_\nu$ is an unobservable quantity. Therefore, a coming question is “Is $\alpha_{ij}$ more sensitive to $f_i^{\rm in}/f_j^{\rm in}$ or to $\delta_i^{q_i}/\delta_j^{q_j}$?” Table\,\ref{table2} and Figure\,\ref{fig-index} show that the average value and distribution of $\alpha_{\rm ox}$ of FSRQs+IBLs are only slightly larger than those of HBLs, even if there is a significant difference in their $\lambda_{\rm Edd}$ (see Figure\,\ref{fig-(V-p)}-(l)). But when the radio emissions are taken into account, remarkable differences are found in the distributions of $\alpha_{\rm ro}$ and $\alpha_{\rm rx}$, which suggests that $\delta_{\rm r}$ plays a larger role than the accretion rates in influencing $\alpha_{ij}$. This implies that the difference in jet properties between FSRQs+LBLs and HBLs may occur in the region from optical to radio (IBLs are considered transitional sources). The jets of HBLs may begin to decelerate in the optical region and exhibit sub-relativistic speeds by the time they propagate to the radio region. The jets of FSRQs+LBLs still maintain relativistic speeds down to the radio region. This analysis agrees with the results that are derived from Figure\,\ref{fig-(V-p)}-(b) (see Section\,\ref{Sec-result}). Therefore, $\delta_{\rm r}/\delta_{\rm x}$ and $\delta_{\rm r}/\delta_{\rm o}$ of FSRQs+LBLs should be larger than those of HBLs. Similar properties were found in radio galaxies; \cite{Giovannini2001ApJ} found that the jets of FR Is and FR IIs are both highly relativistic on the parsec (pc) scale, but there are differences on kiloparsec (kpc) scale. The jets of FR Is may be strongly decelerated within several hundred pc from the core, and often exhibit sub-relativistic speeds and decollimated characteristics as they propagate out to kpc scales \citep{Bhattacharjee2024ApJ}. while the jets of FR IIs show well-collimated characteristics and relativistic speeds up to $\sim100$\,kpc \citep{Boccardi2021A&A,Seo2021ApJ}.

Continuous jet ($q=2+\alpha_\nu$) vs. spherical jet ($q=3+\alpha_\nu$). Historically, the continuous jet refers to the steady jet, while the spherical jet is considered as a sum of discrete features known as blobs or knots, i.e., the so-called intermittent jet \citep[e.g.,][]{Aloy2003ApJ,Fender2004MN,Ferreira2006}, where the compact plasma blobs are inlaid in the collimated outflows \citep{Aloy2003ApJ,Ferreira2006,Vuillaume2018A&A}. BH X-ray binaries (BH-XRBs) are the typical laboratory to study the continuous jets and intermittent jets \citep[e.g.,][]{Fender2004MN,Ferreira2006,Fender2012Sci}. Continuous jets occur in the low/hard state of BH-XRBs, where the accretion disk is usually described by ADAF. With the increase of the accretion rate, BH-XRBs enter the luminous intermediate state, in which the thin disk almost extends down to the innermost stable circular orbit, and the intermittent jets emerge \citep{Fender2004MN,Ferreira2006,Fender2012Sci,Begelman2014ApJL}. The intermittent jet has a larger bulk Lorentz factor \citep[see Figure\,7 in][]{Fender2004MN}. The ultra-relativistic blobs, which are produced when adequate conditions are fulfilled \citep{Ferreira2006}, have been used to explain the highly relativistic phenomena such as super-luminal motions in quasars \citep{Pelletier1989A&A,Pelletier1992MN,Aloy2003ApJ}.

The blobs are formed from the electron-positron ($\rm e^+-e^-$) pairs \citep{Henri1991ApJ,Ferreira2006}. Theoretically, there are two formation channels for $\rm e^+-e^-$ pairs: (1) $\gamma$-ray photons with energies near threshold ($\hbar \omega \gtrsim2m_e c^2$) can annihilate themselves to produce $\rm e^+-e^-$ pairs in superstrong magnetic fields \citep[$\phi_{\rm B}\gtrsim m_e^2c^3/e\hbar$, see][]{Daugherty1983ApJ}, where the squared magnetic fields are proportional to the accretion rates \citep[$\phi_{\rm B}^2\propto \dot{M}$, e.g.,][]{Heinz2003MN,Ghisellini2014Nat,Zamaninasab2014Nat}; (2) the $\gamma-\gamma$ interaction, i.e., $\gamma+\gamma \to e^+ + e^-$, can occur when high-energy $\gamma$-ray photons interact with other photons, including soft photons from the external radiation field (ERF), which originates from the accretion disk, BLR, and dusty torus. The latter process is generally the dominant mechanism, particularly through $\gamma+\gamma_{\rm soft} \to e^+ + e^-$. Continuous generation of $\rm e^+-e^-$ pairs accumulates to form $\rm e^+-e^-$ plasma clouds, ultimately giving rise to blobs \citep{Svensson1987MN,Henri1991ApJ}. Then the blobs are accelerated by radiation pressure through the Compton rocket effect \citep{O-Dell1981ApJ,Ferreira2006,Vuillaume2018A&A}, which requires a locally anisotropic and intense radiation field in the inner zone of the near Eddington accretion disk. Therefore, the ultra-relativistic blobs are usually produced in the jets of high-accretion BH. 

Hence, the jets of FSRQs+LBLs should be intermittent jet ($q=3+\alpha_\nu$) in which the compact blobs move ultra-relativistically in the channel of outflows, while the jets of HBLs should seem to be the continuous jets ($q=2+\alpha_\nu$). Indeed, there were many VLBA studies of pc-scale jets for FSRQs and LBLs \citep[e.g.,][]{Pearson1981Nat,Jorstad2001ApJS,Jorstad2017ApJ,Lister2009AJ,Lister2013AJ,Harwood2022A&A,Yi2024A&A,Shang2025ApJ,Thevenet2026A&A}, showing blobs emanating from a stationary core at apparent super-luminal speeds. This two-flow model has also been applied in FSRQs, the implemented work can be see in \cite{Vuillaume2018A&A}; the three-dimensional, relativistic, hydrodynamic simulation can be see in \cite{Aloy2003ApJ}. However, the weak jets of HBLs assume a plume-like morphology beyond a few milliarcseconds from the core \citep{Piner2008ApJ}. The analyses above imply the $f_\nu^{\rm ob}$ of FSRQs+LBLs is more sensitive to the Doppler beaming effect than that of HBLs.

There are also the differences of spectral indices ($\alpha_{\rm r}$, $\alpha_{\rm o}$, $\alpha_{\rm x}$) among blazars. \cite{Fuhrmann2016A&A} showed that FSRQs have an average value of $\langle \alpha_{\rm r}\rangle=0.02$, BL Lacs have $\langle \alpha_{\rm r}\rangle=0.08$. \cite{Falomo1994ApJS} gave the $\langle \alpha_{\rm o}\rangle=0.75$, $\langle \alpha_{\rm o}\rangle=1.05$, and $\langle \alpha_{\rm o}\rangle=0.65$ for FSRQs, LBLs, and HBLs, respectively. The $\langle \alpha_{\rm x}\rangle$ of FSRQs, LBLs, and HBLs are presented in Section\,\ref{Sec-2.1} (i.e., 0.59, 0.92, and 1.24). But $\delta_i/\delta_j$ appears to make a pivotal role in determining $\alpha_{ij}$ , while the jet shape and $\alpha_\nu$ play a minor role.

\subsection{The Disk-Jet Connection for the Blazar Sequence}

The formation mechanism of relativistic jets is still controversial, but it is widely believed that the large-scale magnetic fields threading accretion disk ($\phi_{\rm B}^2\propto \dot{M}$) play a critical role in jet formation \citep[e.g.,][]{Blandford1977MN,Blandford1982MN,Koide2002Sci,Zamaninasab2014Nat,Yang2024SciA}. Therefore, it can be predicted that a more powerful accretion produces a more powerful jet, and gives rise to a larger Doppler factor. This is consistent with our observational results in Figure\,\ref{fig-(P+Edd+D)}. Figure\,\ref{fig-(P+Edd+D)}-(a) shows a $\log P_{\rm jet}^{\rm in}-\lambda_{\rm Edd}$ correlation for our blazars, suggesting a connection between jet and disk. However, there is a break at $\lambda_{\rm Edd}\sim-2.651$ in the $\log P_{\rm jet}^{\rm in}-\lambda_{\rm Edd}$ correlation, which implies the transition of accretion mode \citep{Ghisellini2011MN,Sbarrato2014MN}. The lower accretion-rate systems (HBLs) are typically described by ADAF, which is characterized by virially hot, geometrically thick, and optically thin \citep[e.g.,][]{Yuan2014ARA&A}. The higher accretion-rate systems (FSRQs+LBLs) are usually well described by the radiatively efficient standard thin disk \citep[][hereafter SSD]{Shakura1973A&A}, which consists of relatively cool, optically thick gas and is geometrically thin and optically thick. Considering the gradual evolution of the accretion disk, IBLs appear to be the transitional objects between FSRQs+LBLs and HBLs. Our results are consistent with \cite{Meyer2011ApJ} and \cite{Keenan2021MN}. The summary of our results can be condensed into Figure\,\ref{fig-sequence}.

\subsection{Discussion of Uncertainties in Our Analysis and Possible Alternative Interpretations}

In this study, the main sources of uncertainty may arise from the representativeness of the sample, the estimation of $\delta_{\rm r}$, the empirical prescription adopted in Eq\,(\ref{Eq-jetmode}), the estimates of $L_{\rm disk}$ and $M_{\rm BH}$ for sources lacking direct measurements, and the extrapolation of $f_{\rm x}$. In addition, inverse-Compton cooling may also have some influence on our results. Therefore, it is necessary to briefly discuss these potential sources of uncertainty.

As shown in Section~\ref{Sec-result}, the results presented in Figures~\ref{fig-(V-p)} and \ref{fig-index} are consistent with those of previous studies, irrespective of the sample size adopted. This suggests that our sample could be regarded as representative of the overall blazar population. However, the number of IBLs in our sample is relatively small, and a larger IBL sample is required to further verify our results. Nevertheless, we believe that our results are reliable because previous studies based on much larger samples have shown that IBLs are distributed nearly equally between the URR and LLR \citep[e.g.,][]{Abdo2010ApJ,Fan2016ApJS,Yang2022ApJS}.

In this work, our $\delta_{\rm r}$ are taken from different studies and they were estimated using different methods \citep[see][]{Hovatta2009,Wu2014,Liodakis2017MN,Liodakis2018ApJ,Ye2021PASJ}. Additionally, $\delta_{\rm r}$ from \cite{Hovatta2009}, \cite{Liodakis2017MN} and \cite{Liodakis2018ApJ} are extrapolated to $\delta_{\rm 5GHz}$ using Eq\,(\ref{Eq-jetmode}). However, we predict this does not have a significant effect on our results; the logical explanations are given as following. It is well-known that the Doppler factor of blazars is notoriously difficult to estimate due to the unobservability of $\beta$ and $\theta$. Although many previous studies have proposed different methods for estimating the Doppler factor \citep[e.g.,][]{Ghisellini1993ApJ,Hovatta2009,Fan2014RAA,Liodakis2017MN,Liodakis2018ApJ,Ye2021PASJ,Pei2022ApJ}, the constraint of the Doppler factor of blazars is still an open question. There are some differences between the $\delta_{\rm r}$ values that are estimated by different methods, but their results are strikingly consistent in showing an anti-correlation between $\delta_{\rm r}$ and $\nu_{\rm p}^{\rm syn}$ \citep[e.g.,][]{Wu2007A&A,Nieppola2008A&A,Yang2022ApJ,Long2025ApJ}, which agrees with our result. For Eq\,(\ref{Eq-jetmode}), we consider it to be applicable for extrapolations between radio bands, as discussed in Section~\ref{Sec-2.1}. Therefore, we believe that our results are robust.

The estimation of $L_{\rm disk}$ and $M_{\rm BH}$ for sources lacking direct measurements has already been discussed in Section~\ref{Sec-2.1} and Appendix~\ref{appendix}, which does not have a significant effect on our results. Most of the $f_{\rm x}$ values used in our analysis are extrapolated from other soft X-ray bands. This also does not have a significant effect on our results. Indeed, many previous studies have adopted the 1\,keV X-ray band (see Section\,\ref{Sec-intro}). However, direct measurements of the 1\,keV X-ray flux are limited, and the corresponding 1\,keV fluxes used in those studies were likewise extrapolated from other soft X-ray bands. Following previous work, we also extrapolated the soft X-ray fluxes to the 1\,keV band. Furthermore, most of the blazars in our sample have available 0.1--2.4\,keV X-ray flux measurements. We verified that using the observed 0.1--2.4\,keV fluxes instead of the extrapolated 1\,keV fluxes yields essentially the same results.

The blazar sequence simply refers to the decrease of $\nu_{\rm p}^{\rm syn}$ with increasing luminosity; in other words, more powerful blazars tend to have redder SED. This phenomenological trend was explained by the difference of accretion power \citep[e.g.,][]{Fossati1998MN,Bottcher2002ApJ,Wang2002ApJ}. However, \cite{Ghisellini1998MN} suggested that blazars with lower $\nu_{\rm p}^{\rm syn}$ have a higher energy density of the ERF, resulting in an increasing amount of Compton cooling. As a result, the electron energy distribution is suppressed at the high-energy end, causing the SED to shift toward lower frequencies. Compton cooling suppresses the high-energy electron population, thereby increasing the relative abundance of low-energy electrons responsible for the radio emission. This may increase $f_{\rm r}^{\rm in}/f_{\rm o}^{\rm in}$ and $f_{\rm r}^{\rm in}/f_{\rm x}^{\rm in}$, thereby providing a possible explanation for our results. However, if strong external-Compton scattering induced by an isotropic ERF dominates, it would result in Compton drag and decelerate the fast jet \citep{G&T2010MN}, which appears to be inconsistent with VLBA radio observations and the $\delta_{\rm r}-\nu_{\rm p}^{\rm syn}$ anti-correlation. Therefore, we suggest that Compton cooling may play a role in shaping the distribution of blazars on the $\alpha_{\rm ro}-\alpha_{\rm ox}$ plane, but it is unlikely to be the dominant factor.

\section{Summary}\label{Sec-summary}

The bimodal distribution of blazars on the $\alpha_{\rm ro}-\alpha_{\rm ox}$ plane has been found in many previous works \citep{Ledden1985ApJ,Padovani1995ApJ,Donato2001A&A,Nieppola2006A&A,Abdo2010ApJ,Fan2016ApJS,Yang2022ApJS}. However, the underlying physics behind the puzzling bimodality has not yet been thoroughly investigated. In this paper, we compile a blazar sample that consists of 136 FSRQs, 64 LBLs, 39 IBLs, and 105 HBLs to explore the possible physics behind this puzzling bimodality. Our main results can be summarized as follows:

(1) We find an anti-correlation of $\langle \delta_{\rm 5GHz}\rangle-\log \nu_{\rm p}^{\rm syn}$, which is consistent with VLBA observations. This at least implies that the jets of HBLs become sub-relativistic in the radio region, whereas the jets in strong-jet sources (FSRQs + LBLs) remain highly relativistic down to radio scales. We also find a positive correlation among $\langle \delta_{\rm 5GHz}\rangle$, $\log P_{\rm jet}^{\rm in}$ , and $\lambda_{\rm Edd}$ for the blazar sequence (i.e., $\langle \delta_{\rm 5GHz}\rangle \propto P_{\rm jet}^{\rm in} \propto \lambda_{\rm Edd}$). This suggests differences in jet strength and accretion power among blazars, where more powerful jets are preferentially produced in systems with higher accretion rates ($\phi_{\rm B}^2\propto \dot{M}$), and such stronger jets are likely to result in larger Doppler factors.

(2) The blazars show a \reflectbox{$\angle$}-shaped evolutionary sequence of (FSRQs+LBLs)$\rightarrow$IBLs$\rightarrow$HBLs on the $\alpha_{\rm ro}-\alpha_{\rm ox}$ plane, with $\langle \delta_{\rm 5GHz}\rangle$, $\log P_{\rm jet}^{\rm in}$, and $\lambda_{\rm Edd}$ decreasing along this \reflectbox{$\angle$}-shaped track. FSRQs+LBLs(63/64) occupy the URR of the plane with $\alpha_{\rm rx}>0.75$, HBLs occupy LLR of the plane with $\alpha_{\rm rx}<0.75$, while IBLs are almost equally distributed URR (19/39) and LLR (20/39). In addition, the distributions of $\alpha_{\rm ox}$ exhibit only slight differences among blazar subclasses, whereas those of $\alpha_{\rm ro}$ and $\alpha_{\rm rx}$ differ significantly. Our results can be naturally explained within a framework in which the distribution of blazars on the $\alpha_{\rm ro}-\alpha_{\rm ox}$ plane is regulated by the accretion rates and the Doppler beaming effect. However, the distribution of blazars appears to be more sensitive to the beaming effect than to accretion rates. Further, we find that the difference of beaming effect in different bands ($\delta_i/\delta_j$) plays a pivotal role in the distribution of blazars, the jet shape ($q=2(3)+\alpha_\nu$) and spectral indices ($\alpha_\nu$) play a minor role. Our results imply that the main differences of jet properties between FSRQs+LBLs and HBLs occur in the region from optical to radio. The jets of HBLs may begin to decelerate in the optical region and exhibit sub-relativistic speeds as they propagate to the radio region; the jets of FSRQs+LBLs are accelerated to a larger distance from the core and still maintain relativistic speeds down to the radio region. IBLs are considered transitional objects between them.

In addition, Compton cooling may play a role in shaping the distribution of blazars on the $\alpha_{\rm ro}-\alpha_{\rm ox}$ plane, but it is unlikely to be the dominant mechanism. Further observational and theoretical studies are needed to verify this interpretation.

(3) Our results show that, in the blazar sequence, there is a significant distinction of physical properties between FSRQs+LBLs and HBLs, while IBLs appear to serve as a bridge between FSRQs+LBLs and HBLs. On average, FSRQs+LBLs are the strong-jet sources coupled with SSD, HBLs are the weak-jet sources coupled with ADAF, while IBLs are considered to be transitional objects.

\begin{acknowledgments}
I thank Prof. Junhui Fan (my mentor) and the anonymous referee for the constructive and useful comments and suggestions, which have helped me to improve this paper. This work is partially supported by the National Natural Science Foundation of China (NSFC grants 12433004, 12133004 and 12303019), the National Key Research and Development Program of China (Grant No. 2025YFA1614102), the Eighteenth Regular Meeting Exchange Project of the Scientific and Technological Cooperation Committee between the People’s Republic of China and the Republic of Bulgaria (Series No. 1802), grants from the China Manned Space Project (Grant No. CMS-CSST-2025-A07). Additional support was provided for the Astrophysics Key Disciplines of Guangdong Province and Guangzhou City, as well as the Guangdong Major Project of Basic and Applied Basic Research (grant No. 2024A1515013169).
\end{acknowledgments}

\appendix

\section{The estimate of $L_{\rm disk}$ and $M_{\rm BH}$}\label{appendix}

For simplicity, we present the composition of our sample in Table\,\ref{table3}, which facilitates further analysis. There are 127/136 FSRQs, 34/64 LBLs, 8/39 IBLs, and 42/105 HBLs that have both available $L_{\rm disk}$ and $M_{\rm BH}$. If sources without available $L_{\rm disk}$ or $M_{\rm BH}$ are eliminated, the substantial reduction of BL Lacs severely affects the statistical results, especially IBLs. For example, only 8 IBLs have available $L_{\rm disk}$ and $M_{\rm BH}$, with 6 IBLs occupying the LLR and 2 IBLs occupying the URR. However, previous works suggested that IBLs are the transitional objects between LBLs and HBLs \citep[e.g.,][]{Nieppola2006A&A,Abdo2010ApJ}. Therefore, for the sources without available $L_{\rm disk}$ or $M_{\rm BH}$, we try to investigate a correlation between $L_{\rm disk}$ and other luminosities to estimate their $L_{\rm disk}$, and use the FP to estimate their $M_{\rm BH}$. 

Because the broad-band emissions from AGNs are proportional to accretion rates ($f_\nu^{\rm in}\propto \dot{M}^{\xi_\nu}$: see Section\,\ref{Sec-4.1}), we predict that there could be a correlation between $L_{\rm disk}$ and $L_\nu$. We find that there is a strong correlation between $L_{\rm disk}$ and $L_{\rm x}^{\rm ob}$ for our blazars. Similarly, we divide our sample into FSRQs+LBLs+$\rm IBL_{>0.75}$ and HBLs+$\rm IBL_{<0.75}$. Using the ordinary least-square (OLS) bisector method \citep{Isobe1990ApJ}, we obtain the $L_{\rm disk}-L_{\rm x}^{\rm ob}$ relation of FSRQs+LBLs+$\rm IBL_{>0.75}$ with available $L_{\rm disk}$ as:

\begin{equation}
    \log L_{\rm disk}=(1.238\pm0.050)\log L_{\rm x}^{\rm ob}-(9.927\pm2.229)
    \label{Eq-A1}
\end{equation}
with a scatter of $\sigma_L=0.626$\,dex, and a Spearman correlation coefficient of $R=0.749$ and $P=7.639\times10^{-32}$. The $L_{\rm disk}-L_{\rm x}^{\rm ob}$ relation of HBLs+$\rm IBL_{<0.75}$ with available $L_{\rm disk}$ is:

\begin{equation}
    \log L_{\rm disk}=(0.901\pm0.098)\log L_{\rm x}^{\rm ob}-(3.820\pm4.304)
    \label{Eq-A2}
\end{equation}
with a scatter of $\sigma_L=0.591$\,dex, and a Spearman correlation coefficient of $R=0.737$ and $P=4.155\times10^{-11}$. 

For the sources without available $M_{\rm BH}$, we use FP to estimate their $M_{\rm BH}$. The FP is a relation between radio luminosity, X-ray luminosity, and BH mass, which was firstly found by \cite{Merloni2003MN}. Subsequently, FP was developed to estimate BH mass \citep[e.g.,][]{Gultekin2009ApJ,Bariuan2022MN}. \cite{Long2025ApJ,Long2026MN} found that FSRQs+LBLs and HBLs follow different scaling relations of FP. Following \cite{Gultekin2009ApJ} and \cite{Bariuan2022MN}, we define the FP as:

\begin{equation}
    \log M_{\rm BH}=\xi_{\rm mr}\log L_{\rm r}^{\rm ob}+\xi_{\rm mx}\log L_{\rm x}^{\rm ob}+c_{0}
\end{equation}
By using the OLS bisector method, we obtain the FP of FSRQs+LBLs+$\rm IBL_{>0.75}$ with available $M_{\rm BH}$ as:

\begin{equation}
    \log M_{\rm BH}=(0.235\pm0.116)\log L_{\rm r}^{\rm ob}+(0.334\pm0.140)\log L_{\rm x}^{\rm ob}-16.558
    \label{Eq-A4}
\end{equation}
with a scatter of $\sigma_{\rm M}=0.464$\,dex. The FP of HBLs+$\rm IBL_{<0.75}$ with available $M_{\rm BH}$ is:

\begin{equation}
    \log M_{\rm BH}=(0.422\pm0.315)\log L_{\rm r}^{\rm ob}+(0.009\pm0.202)\log L_{\rm x}^{\rm ob}-8.941
    \label{Eq-A5}
\end{equation}
with a scatter of $\sigma_{\rm M}=0.543$\,dex.

These methods do not have a significant effect on our results. Actually, similar methods were also used to estimate $L_{\rm disk}$ and $M_{\rm BH}$ for blazars \citep[e.g.,][]{D-Elia2003MN,Sbarrato2012MN,Sbarrato2014MN,Sheng2024ApJ}. Additionally, the $\lambda_{\rm Edd}-\log \nu_{\rm p}^{\rm syn}$ relation of our sample is consistent with that of \cite{Paliya2021ApJS}, which implies that our results do not have a significant bias.

\begin{deluxetable*}{ccccc}
\tablecaption{The Composition of Our Sample\label{table3}}
		\renewcommand{\tabcolsep}{6pt}
    \tablehead{ Sample & FSRQ & LBL & IBL & HBL }

	\startdata
	Total & 136 & 64 & 39 & 105 \\
    $L_{\rm disk}$ & 130 & 36 & 14 & 48 \\
	$M_{\rm BH}$ & 130 & 48 & 20 & 83 \\
    $L_{\rm disk}+M_{\rm BH}$ & 127 & 34 & 8 & 42 \\
    \enddata
    
    \tablecomments{Total denotes the entire sample. $L_{\rm disk}$ denotes the subsample that consists of blazars with available $L_{\rm disk}$. $M_{\rm BH}$ denotes the subsample that consists of blazars with available $M_{\rm BH}$. $L_{\rm disk}+M_{\rm BH}$ denotes the subsample that consists of blazars with both available $L_{\rm disk}$ and $M_{\rm BH}$.}
\end{deluxetable*}


\bibliography{sample701}{}

@ARTICLE{Urry&Padovani1995PASP,
       author = {{Urry}, C. Megan and {Padovani}, Paolo},
        title = "{Unified Schemes for Radio-Loud Active Galactic Nuclei}",
      journal = {\pasp},
         year = 1995,
        month = sep,
       volume = {107},
        pages = {803},
          doi = {10.1086/133630},
archivePrefix = {arXiv},
       eprint = {astro-ph/9506063},
 primaryClass = {astro-ph},
       adsurl = {https://ui.adsabs.harvard.edu/abs/1995PASP..107..803U}
}

@ARTICLE{Wills1992ApJ,
       author = {{Wills}, Beverley J. and {Wills}, D. and {Breger}, Michel and {Antonucci}, R.~R.~J. and {Barvainis}, Richard},
        title = "{A Survey for High Optical Polarization in Quasars with Core-dominant Radio Structure: Is There a Beamed Optical Continuum?}",
      journal = {\apj},
         year = 1992,
        month = oct,
       volume = {398},
        pages = {454},
          doi = {10.1086/171869},
       adsurl = {https://ui.adsabs.harvard.edu/abs/1992ApJ...398..454W}
}

@ARTICLE{Fan1996A&A,
       author = {{Fan}, J.~H. and {Xie}, G.~Z.},
        title = "{The properties of BL Lacertae objects.}",
      journal = {\aap},
         year = 1996,
        month = feb,
       volume = {306},
        pages = {55},
       adsurl = {https://ui.adsabs.harvard.edu/abs/1996A&A...306...55F}
}

@ARTICLE{Fan2000ApJ,
       author = {{Fan}, J.~H. and {Lin}, R.~G.},
        title = "{Optical Variability and Periodicity Analysis for Blazars. I. Light Curves for Radio-selected BL Lacertae Objects}",
      journal = {\apj},
         year = 2000,
        month = jul,
       volume = {537},
       number = {1},
        pages = {101-122},
          doi = {10.1086/308996},
       adsurl = {https://ui.adsabs.harvard.edu/abs/2000ApJ...537..101F}
}

@ARTICLE{Fan2005ChJAS,
       author = {{Fan}, Jun-Hui},
        title = "{Optical Variability of Blazars}",
      journal = {ChJAS},
         year = 2005,
        month = jun,
       volume = {5},
        pages = {213-223},
       adsurl = {https://ui.adsabs.harvard.edu/abs/2005ChJAS...5..213F}
}

@ARTICLE{Fan2013RAA,
       author = {{Fan}, Jun-Hui and {Yang}, Jiang-He and {Liu}, Yi and {Zhang}, Jing-Yi},
        title = "{The gamma-ray Doppler factor determinations for a Fermi blazar sample}",
      journal = {RAA},
         year = 2013,
        month = mar,
       volume = {13},
       number = {3},
        pages = {259-269},
          doi = {10.1088/1674-4527/13/3/002},
       adsurl = {https://ui.adsabs.harvard.edu/abs/2013RAA....13..259F}
}

@ARTICLE{Raiteri2017Natur,
       author = {{Raiteri}, C.~M. and {Villata}, M. and {Acosta-Pulido}, J.~A. and {Agudo}, I. and {Arkharov}, A.~A. and {Bachev}, R. and {Baida}, G.~V. and {Ben{\'\i}tez}, E. and {Borman}, G.~A. and {Boschin}, W. and {Bozhilov}, V. and {Butuzova}, M.~S. and {Calcidese}, P. and {Carnerero}, M.~I. and {Carosati}, D. and {Casadio}, C. and {Castro-Segura}, N. and {Chen}, W.-P. and {Damljanovic}, G. and {D'Ammando}, F. and {di Paola}, A. and {Echevarr{\'\i}a}, J. and {Efimova}, N.~V. and {Ehgamberdiev}, Sh. A. and {Espinosa}, C. and {Fuentes}, A. and {Giunta}, A. and {G{\'o}mez}, J.~L. and {Grishina}, T.~S. and {Gurwell}, M.~A. and {Hiriart}, D. and {Jermak}, H. and {Jordan}, B. and {Jorstad}, S.~G. and {Joshi}, M. and {Kopatskaya}, E.~N. and {Kuratov}, K. and {Kurtanidze}, O.~M. and {Kurtanidze}, S.~O. and {L{\"a}hteenm{\"a}ki}, A. and {Larionov}, V.~M. and {Larionova}, E.~G. and {Larionova}, L.~V. and {L{\'a}zaro}, C. and {Lin}, C.~S. and {Malmrose}, M.~P. and {Marscher}, A.~P. and {Matsumoto}, K. and {McBreen}, B. and {Michel}, R. and {Mihov}, B. and {Minev}, M. and {Mirzaqulov}, D.~O. and {Mokrushina}, A.~A. and {Molina}, S.~N. and {Moody}, J.~W. and {Morozova}, D.~A. and {Nazarov}, S.~V. and {Nikolashvili}, M.~G. and {Ohlert}, J.~M. and {Okhmat}, D.~N. and {Ovcharov}, E. and {Pinna}, F. and {Polakis}, T.~A. and {Protasio}, C. and {Pursimo}, T. and {Redondo-Lorenzo}, F.~J. and {Rizzi}, N. and {Rodriguez-Coira}, G. and {Sadakane}, K. and {Sadun}, A.~C. and {Samal}, M.~R. and {Savchenko}, S.~S. and {Semkov}, E. and {Skiff}, B.~A. and {Slavcheva-Mihova}, L. and {Smith}, P.~S. and {Steele}, I.~A. and {Strigachev}, A. and {Tammi}, J. and {Thum}, C. and {Tornikoski}, M. and {Troitskaya}, Yu. V. and {Troitsky}, I.~S. and {Vasilyev}, A.~A. and {Vince}, O.},
        title = "{Blazar spectral variability as explained by a twisted inhomogeneous jet}",
      journal = {\nat},
         year = 2017,
        month = dec,
       volume = {552},
       number = {7685},
        pages = {374-377},
          doi = {10.1038/nature24623},
archivePrefix = {arXiv},
       eprint = {1712.02098},
 primaryClass = {astro-ph.HE},
       adsurl = {https://ui.adsabs.harvard.edu/abs/2017Natur.552..374R}
}

@ARTICLE{Abdollahi2022ApJS,
       author = {{Abdollahi}, S. and {Acero}, F. and {Baldini}, L. and {Ballet}, J. and {Bastieri}, D. and {Bellazzini}, R. and {Berenji}, B. and {Berretta}, A. and {Bissaldi}, E. and {Blandford}, R.~D. and {Bloom}, E. and {Bonino}, R. and {Brill}, A. and {Britto}, R.~J. and {Bruel}, P. and {Burnett}, T.~H. and {Buson}, S. and {Cameron}, R.~A. and {Caputo}, R. and {Caraveo}, P.~A. and {Castro}, D. and {Chaty}, S. and {Cheung}, C.~C. and {Chiaro}, G. and {Cibrario}, N. and {Ciprini}, S. and {Coronado-Bl{\'a}zquez}, J. and {Crnogorcevic}, M. and {Cutini}, S. and {D'Ammando}, F. and {De Gaetano}, S. and {Digel}, S.~W. and {Di Lalla}, N. and {Dirirsa}, F. and {Di Venere}, L. and {Dom{\'\i}nguez}, A. and {Fallah Ramazani}, V. and {Fegan}, S.~J. and {Ferrara}, E.~C. and {Fiori}, A. and {Fleischhack}, H. and {Franckowiak}, A. and {Fukazawa}, Y. and {Funk}, S. and {Fusco}, P. and {Galanti}, G. and {Gammaldi}, V. and {Gargano}, F. and {Garrappa}, S. and {Gasparrini}, D. and {Giacchino}, F. and {Giglietto}, N. and {Giordano}, F. and {Giroletti}, M. and {Glanzman}, T. and {Green}, D. and {Grenier}, I.~A. and {Grondin}, M.-H. and {Guillemot}, L. and {Guiriec}, S. and {Gustafsson}, M. and {Harding}, A.~K. and {Hays}, E. and {Hewitt}, J.~W. and {Horan}, D. and {Hou}, X. and {J{\'o}hannesson}, G. and {Karwin}, C. and {Kayanoki}, T. and {Kerr}, M. and {Kuss}, M. and {Landriu}, D. and {Larsson}, S. and {Latronico}, L. and {Lemoine-Goumard}, M. and {Li}, J. and {Liodakis}, I. and {Longo}, F. and {Loparco}, F. and {Lott}, B. and {Lubrano}, P. and {Maldera}, S. and {Malyshev}, D. and {Manfreda}, A. and {Mart{\'\i}-Devesa}, G. and {Mazziotta}, M.~N. and {Mereu}, I. and {Meyer}, M. and {Michelson}, P.~F. and {Mirabal}, N. and {Mitthumsiri}, W. and {Mizuno}, T. and {Moiseev}, A.~A. and {Monzani}, M.~E. and {Morselli}, A. and {Moskalenko}, I.~V. and {Negro}, M. and {Nuss}, E. and {Omodei}, N. and {Orienti}, M. and {Orlando}, E. and {Paneque}, D. and {Pei}, Z. and {Perkins}, J.~S. and {Persic}, M. and {Pesce-Rollins}, M. and {Petrosian}, V. and {Pillera}, R. and {Poon}, H. and {Porter}, T.~A. and {Principe}, G. and {Rain{\`o}}, S. and {Rando}, R. and {Rani}, B. and {Razzano}, M. and {Razzaque}, S. and {Reimer}, A. and {Reimer}, O. and {Reposeur}, T. and {S{\'a}nchez-Conde}, M. and {Saz Parkinson}, P.~M. and {Scotton}, L. and {Serini}, D. and {Sgr{\`o}}, C. and {Siskind}, E.~J. and {Smith}, D.~A. and {Spandre}, G. and {Spinelli}, P. and {Sueoka}, K. and {Suson}, D.~J. and {Tajima}, H. and {Tak}, D. and {Thayer}, J.~B. and {Thompson}, D.~J. and {Torres}, D.~F. and {Troja}, E. and {Valverde}, J. and {Wood}, K. and {Zaharijas}, G.},
        title = "{Incremental Fermi Large Area Telescope Fourth Source Catalog}",
      journal = {\apjs},
         year = 2022,
        month = jun,
       volume = {260},
       number = {2},
          eid = {53},
        pages = {53},
          doi = {10.3847/1538-4365/ac6751},
archivePrefix = {arXiv},
       eprint = {2201.11184},
 primaryClass = {astro-ph.HE},
       adsurl = {https://ui.adsabs.harvard.edu/abs/2022ApJS..260...53A}
}

@ARTICLE{Scarpa1997A&A,
       author = {{Scarpa}, R. and {Falomo}, R.},
        title = "{Are high polarization quasars and BL Lacertae objects really different? A study of the optical spectral properties.}",
      journal = {\aap},
         year = 1997,
        month = sep,
       volume = {325},
        pages = {109-123},
       adsurl = {https://ui.adsabs.harvard.edu/abs/1997A&A...325..109S}
}

@ARTICLE{Corbett2000MN,
       author = {{Corbett}, E.~A. and {Robinson}, A. and {Axon}, D.~J. and {Hough}, J.~H.},
        title = "{A Seyfert-like nucleus concealed in BL Lacertae?}",
      journal = {\mnras},
         year = 2000,
        month = jan,
       volume = {311},
       number = {3},
        pages = {485-492},
          doi = {10.1046/j.1365-8711.2000.03045.x},
       adsurl = {https://ui.adsabs.harvard.edu/abs/2000MNRAS.311..485C}
}

@ARTICLE{Fan2003ApJL,
       author = {{Fan}, J.~H.},
        title = "{Relation between BL Lacertae Objects and Flat-Spectrum Radio Quasars}",
      journal = {\apjl},
         year = 2003,
        month = mar,
       volume = {585},
       number = {1},
        pages = {L23-L24},
          doi = {10.1086/374033},
       adsurl = {https://ui.adsabs.harvard.edu/abs/2003ApJ...585L..23F}
}

@ARTICLE{Ghisellini2008MN,
       author = {{Ghisellini}, G. and {Tavecchio}, F.},
        title = "{The blazar sequence: a new perspective}",
      journal = {\mnras},
         year = 2008,
        month = jul,
       volume = {387},
       number = {4},
        pages = {1669-1680},
          doi = {10.1111/j.1365-2966.2008.13360.x},
archivePrefix = {arXiv},
       eprint = {0802.1918},
 primaryClass = {astro-ph},
       adsurl = {https://ui.adsabs.harvard.edu/abs/2008MNRAS.387.1669G}
}

@ARTICLE{Ghisellini2011MN,
       author = {{Ghisellini}, G. and {Tavecchio}, F. and {Foschini}, L. and {Ghirlanda}, G.},
        title = "{The transition between BL Lac objects and flat spectrum radio quasars}",
      journal = {\mnras},
         year = 2011,
        month = jul,
       volume = {414},
       number = {3},
        pages = {2674-2689},
          doi = {10.1111/j.1365-2966.2011.18578.x},
archivePrefix = {arXiv},
       eprint = {1012.0308},
 primaryClass = {astro-ph.CO},
       adsurl = {https://ui.adsabs.harvard.edu/abs/2011MNRAS.414.2674G}
}

@ARTICLE{Padovani2012MN,
       author = {{Padovani}, P. and {Giommi}, P. and {Rau}, A.},
        title = "{The discovery of high-power high synchrotron peak blazars}",
      journal = {\mnras},
         year = 2012,
        month = may,
       volume = {422},
       number = {1},
        pages = {L48-L52},
          doi = {10.1111/j.1745-3933.2012.01234.x},
archivePrefix = {arXiv},
       eprint = {1202.2236},
 primaryClass = {astro-ph.CO},
       adsurl = {https://ui.adsabs.harvard.edu/abs/2012MNRAS.422L..48P}
}

@ARTICLE{Foschini2012RAA,
       author = {{Foschini}, Luigi},
        title = "{On the emission lines in active galactic nuclei with relativistic jets}",
      journal = {RAA},
         year = 2012,
        month = apr,
       volume = {12},
       number = {4},
        pages = {359-368},
          doi = {10.1088/1674-4527/12/4/001},
archivePrefix = {arXiv},
       eprint = {1103.2008},
 primaryClass = {astro-ph.GA},
       adsurl = {https://ui.adsabs.harvard.edu/abs/2012RAA....12..359F}
}

@ARTICLE{Sbarrato2014MN,
       author = {{Sbarrato}, T. and {Padovani}, P. and {Ghisellini}, G.},
        title = "{The jet-disc connection in AGN}",
      journal = {\mnras},
         year = 2014,
        month = nov,
       volume = {445},
       number = {1},
        pages = {81-92},
          doi = {10.1093/mnras/stu1759},
archivePrefix = {arXiv},
       eprint = {1405.4865},
 primaryClass = {astro-ph.HE},
       adsurl = {https://ui.adsabs.harvard.edu/abs/2014MNRAS.445...81S}
}

@ARTICLE{Chen2025MN,
       author = {{Chen}, Guohai and {Yang}, Wenxin and {Liu}, Yi and {Ho}, Luis C. and {Xiao}, Hubing and {Bachev}, Rumen S. and {Strigachev}, Anton and {Fan}, Junhui},
        title = "{Changing-look behaviour and amplitude-modulated QPO in blazar Ton 599}",
      journal = {\mnras},
         year = 2025,
        month = dec,
       volume = {544},
       number = {2},
        pages = {1926-1938},
          doi = {10.1093/mnras/staf1818},
       adsurl = {https://ui.adsabs.harvard.edu/abs/2025MNRAS.544.1926C}
}

@ARTICLE{Fossati1998MN,
       author = {{Fossati}, G. and {Maraschi}, L. and {Celotti}, A. and {Comastri}, A. and {Ghisellini}, G.},
        title = "{A unifying view of the spectral energy distributions of blazars}",
      journal = {\mnras},
         year = 1998,
        month = sep,
       volume = {299},
       number = {2},
        pages = {433-448},
          doi = {10.1046/j.1365-8711.1998.01828.x},
archivePrefix = {arXiv},
       eprint = {astro-ph/9804103},
 primaryClass = {astro-ph},
       adsurl = {https://ui.adsabs.harvard.edu/abs/1998MNRAS.299..433F}
}

@ARTICLE{Donato2001A&A,
       author = {{Donato}, D. and {Ghisellini}, G. and {Tagliaferri}, G. and {Fossati}, G.},
        title = "{Hard X-ray properties of blazars}",
      journal = {\aap},
         year = 2001,
        month = sep,
       volume = {375},
        pages = {739-751},
          doi = {10.1051/0004-6361:20010675},
archivePrefix = {arXiv},
       eprint = {astro-ph/0105203},
 primaryClass = {astro-ph},
       adsurl = {https://ui.adsabs.harvard.edu/abs/2001A&A...375..739D}
}

@ARTICLE{Fan2016ApJS,
       author = {{Fan}, J.~H. and {Yang}, J.~H. and {Liu}, Y. and {Luo}, G.~Y. and {Lin}, C. and {Yuan}, Y.~H. and {Xiao}, H.~B. and {Zhou}, A.~Y. and {Hua}, T.~X. and {Pei}, Z.~Y.},
        title = "{The Spectral Energy Distributions of Fermi Blazars}",
      journal = {\apjs},
         year = 2016,
        month = oct,
       volume = {226},
       number = {2},
          eid = {20},
        pages = {20},
          doi = {10.3847/0067-0049/226/2/20},
archivePrefix = {arXiv},
       eprint = {1608.03958},
 primaryClass = {astro-ph.HE},
       adsurl = {https://ui.adsabs.harvard.edu/abs/2016ApJS..226...20F}
}

@ARTICLE{Blandford2019,
       author = {{Blandford}, Roger and {Meier}, David and {Readhead}, Anthony},
        title = "{Relativistic Jets from Active Galactic Nuclei}",
      journal = {\araa},
         year = 2019,
        month = aug,
       volume = {57},
        pages = {467-509},
          doi = {10.1146/annurev-astro-081817-051948},
archivePrefix = {arXiv},
       eprint = {1812.06025},
 primaryClass = {astro-ph.HE},
       adsurl = {https://ui.adsabs.harvard.edu/abs/2019ARA&A..57..467B}
}

@ARTICLE{Yang2022ApJS,
       author = {{Yang}, J.~H. and {Fan}, J.~H. and {Liu}, Y. and {Tuo}, M.~X. and {Pei}, Z.~Y. and {Yang}, W.~X. and {Yuan}, Y.~H. and {He}, S.~L. and {Wang}, S.~H. and {Wang}, X.~C. and {Chen}, X.~J. and {Qu}, X.~H. and {Cao}, Q. and {Tao}, Q.~Y. and {Zhang}, Y.~L. and {Liu}, C.~Q. and {Nie}, J.~J. and {Liu}, L.~F. and {Jiang}, D.~K. and {Jiang}, A.~N. and {Liu}, B. and {Yang}, R.~S.},
        title = "{The Spectral Energy Distributions for 4FGL Blazars}",
      journal = {\apjs},
         year = 2022,
        month = sep,
       volume = {262},
       number = {1},
          eid = {18},
        pages = {18},
          doi = {10.3847/1538-4365/ac7deb},
       adsurl = {https://ui.adsabs.harvard.edu/abs/2022ApJS..262...18Y}
}

@ARTICLE{Nieppola2006A&A,
       author = {{Nieppola}, E. and {Tornikoski}, M. and {Valtaoja}, E.},
        title = "{Spectral energy distributions of a large sample of BL Lacertae objects}",
      journal = {\aap},
         year = 2006,
        month = jan,
       volume = {445},
       number = {2},
        pages = {441-450},
          doi = {10.1051/0004-6361:20053316},
archivePrefix = {arXiv},
       eprint = {astro-ph/0509045},
 primaryClass = {astro-ph},
       adsurl = {https://ui.adsabs.harvard.edu/abs/2006A&A...445..441N}
}

@ARTICLE{Bottcher2013ApJ,
       author = {{B{\"o}ttcher}, M. and {Reimer}, A. and {Sweeney}, K. and {Prakash}, A.},
        title = "{Leptonic and Hadronic Modeling of Fermi-detected Blazars}",
      journal = {\apj},
         year = 2013,
        month = may,
       volume = {768},
       number = {1},
          eid = {54},
        pages = {54},
          doi = {10.1088/0004-637X/768/1/54},
archivePrefix = {arXiv},
       eprint = {1304.0605},
 primaryClass = {astro-ph.HE},
       adsurl = {https://ui.adsabs.harvard.edu/abs/2013ApJ...768...54B}
}

@ARTICLE{Cerruti2015MN,
       author = {{Cerruti}, M. and {Zech}, A. and {Boisson}, C. and {Inoue}, S.},
        title = "{A hadronic origin for ultra-high-frequency-peaked BL Lac objects}",
      journal = {\mnras},
         year = 2015,
        month = mar,
       volume = {448},
       number = {1},
        pages = {910-927},
          doi = {10.1093/mnras/stu2691},
archivePrefix = {arXiv},
       eprint = {1411.5968},
 primaryClass = {astro-ph.HE},
       adsurl = {https://ui.adsabs.harvard.edu/abs/2015MNRAS.448..910C}
}

@ARTICLE{Abdo2010ApJ,
       author = {{Abdo}, A.~A. and {Ackermann}, M. and {Agudo}, I. and {Ajello}, M. and {Aller}, H.~D. and {Aller}, M.~F. and {Angelakis}, E. and {Arkharov}, A.~A. and {Axelsson}, M. and {Bach}, U. and {Baldini}, L. and {Ballet}, J. and {Barbiellini}, G. and {Bastieri}, D. and {Baughman}, B.~M. and {Bechtol}, K. and {Bellazzini}, R. and {Benitez}, E. and {Berdyugin}, A. and {Berenji}, B. and {Blandford}, R.~D. and {Bloom}, E.~D. and {Boettcher}, M. and {Bonamente}, E. and {Borgland}, A.~W. and {Bregeon}, J. and {Brez}, A. and {Brigida}, M. and {Bruel}, P. and {Burnett}, T.~H. and {Burrows}, D. and {Buson}, S. and {Caliandro}, G.~A. and {Calzoletti}, L. and {Cameron}, R.~A. and {Capalbi}, M. and {Caraveo}, P.~A. and {Carosati}, D. and {Casandjian}, J.~M. and {Cavazzuti}, E. and {Cecchi}, C. and {{\c{C}}elik}, {\"O}. and {Charles}, E. and {Chaty}, S. and {Chekhtman}, A. and {Chen}, W.~P. and {Chiang}, J. and {Chincarini}, G. and {Ciprini}, S. and {Claus}, R. and {Cohen-Tanugi}, J. and {Colafrancesco}, S. and {Cominsky}, L.~R. and {Conrad}, J. and {Costamante}, L. and {Cutini}, S. and {D'ammando}, F. and {Deitrick}, R. and {D'Elia}, V. and {Dermer}, C.~D. and {de Angelis}, A. and {de Palma}, F. and {Digel}, S.~W. and {Donnarumma}, I. and {Silva}, E. do Couto e. and {Drell}, P.~S. and {Dubois}, R. and {Dultzin}, D. and {Dumora}, D. and {Falcone}, A. and {Farnier}, C. and {Favuzzi}, C. and {Fegan}, S.~J. and {Focke}, W.~B. and {Forn{\'e}}, E. and {Fortin}, P. and {Frailis}, M. and {Fuhrmann}, L. and {Fukazawa}, Y. and {Funk}, S. and {Fusco}, P. and {G{\'o}mez}, J.~L. and {Gargano}, F. and {Gasparrini}, D. and {Gehrels}, N. and {Germani}, S. and {Giebels}, B. and {Giglietto}, N. and {Giommi}, P. and {Giordano}, F. and {Giuliani}, A. and {Glanzman}, T. and {Godfrey}, G. and {Grenier}, I.~A. and {Gronwall}, C. and {Grove}, J.~E. and {Guillemot}, L. and {Guiriec}, S. and {Gurwell}, M.~A. and {Hadasch}, D. and {Hanabata}, Y. and {Harding}, A.~K. and {Hayashida}, M. and {Hays}, E. and {Healey}, S.~E. and {Heidt}, J. and {Hiriart}, D. and {Horan}, D. and {Hoversten}, E.~A. and {Hughes}, R.~E. and {Itoh}, R. and {Jackson}, M.~S. and {J{\'o}hannesson}, G. and {Johnson}, A.~S. and {Johnson}, W.~N. and {Jorstad}, S.~G. and {Kadler}, M. and {Kamae}, T. and {Katagiri}, H. and {Kataoka}, J. and {Kawai}, N. and {Kennea}, J. and {Kerr}, M. and {Kimeridze}, G. and {Kn{\"o}dlseder}, J. and {Kocian}, M.~L. and {Kopatskaya}, E.~N. and {Koptelova}, E. and {Konstantinova}, T.~S. and {Kovalev}, Y.~Y. and {Kovalev}, Yu. A. and {Kurtanidze}, O.~M. and {Kuss}, M. and {Lande}, J. and {Larionov}, V.~M. and {Latronico}, L. and {Leto}, P. and {Lindfors}, E. and {Longo}, F. and {Loparco}, F. and {Lott}, B. and {Lovellette}, M.~N. and {Lubrano}, P. and {Madejski}, G.~M. and {Makeev}, A. and {Marchegiani}, P. and {Marscher}, A.~P. and {Marshall}, F. and {Max-Moerbeck}, W. and {Mazziotta}, M.~N. and {McConville}, W. and {McEnery}, J.~E. and {Meurer}, C. and {Michelson}, P.~F. and {Mitthumsiri}, W. and {Mizuno}, T. and {Moiseev}, A.~A. and {Monte}, C. and {Monzani}, M.~E. and {Morselli}, A. and {Moskalenko}, I.~V. and {Murgia}, S. and {Nestoras}, I. and {Nilsson}, K. and {Nizhelsky}, N.~A. and {Nolan}, P.~L. and {Norris}, J.~P. and {Nuss}, E. and {Ohsugi}, T. and {Ojha}, R. and {Omodei}, N. and {Orlando}, E. and {Ormes}, J.~F. and {Osborne}, J. and {Ozaki}, M. and {Pacciani}, L. and {Padovani}, P. and {Pagani}, C. and {Page}, K. and {Paneque}, D. and {Panetta}, J.~H. and {Parent}, D. and {Pasanen}, M. and {Pavlidou}, V. and {Pelassa}, V. and {Pepe}, M. and {Perri}, M. and {Pesce-Rollins}, M. and {Piranomonte}, S. and {Piron}, F. and {Pittori}, C. and {Porter}, T.~A. and {Puccetti}, S. and {Rahoui}, F. and {Rain{\`o}}, S. and {Raiteri}, C. and {Rando}, R. and {Razzano}, M. and {Reimer}, A. and {Reimer}, O.},
        title = "{The Spectral Energy Distribution of Fermi Bright Blazars}",
      journal = {\apj},
         year = 2010,
        month = jun,
       volume = {716},
       number = {1},
        pages = {30-70},
          doi = {10.1088/0004-637X/716/1/30},
archivePrefix = {arXiv},
       eprint = {0912.2040},
 primaryClass = {astro-ph.CO},
       adsurl = {https://ui.adsabs.harvard.edu/abs/2010ApJ...716...30A}
}

@ARTICLE{Giommi2012A&A,
       author = {{Giommi}, P. and {Polenta}, G. and {L{\"a}hteenm{\"a}ki}, A. and {Thompson}, D.~J. and {Capalbi}, M. and {Cutini}, S. and {Gasparrini}, D. and {Gonz{\'a}lez-Nuevo}, J. and {Le{\'o}n-Tavares}, J. and {L{\'o}pez-Caniego}, M. and {Mazziotta}, M.~N. and {Monte}, C. and {Perri}, M. and {Rain{\`o}}, S. and {Tosti}, G. and {Tramacere}, A. and {Verrecchia}, F. and {Aller}, H.~D. and {Aller}, M.~F. and {Angelakis}, E. and {Bastieri}, D. and {Berdyugin}, A. and {Bonaldi}, A. and {Bonavera}, L. and {Burigana}, C. and {Burrows}, D.~N. and {Buson}, S. and {Cavazzuti}, E. and {Chincarini}, G. and {Colafrancesco}, S. and {Costamante}, L. and {Cuttaia}, F. and {D'Ammando}, F. and {de Zotti}, G. and {Frailis}, M. and {Fuhrmann}, L. and {Galeotta}, S. and {Gargano}, F. and {Gehrels}, N. and {Giglietto}, N. and {Giordano}, F. and {Giroletti}, M. and {Keih{\"a}nen}, E. and {King}, O. and {Krichbaum}, T.~P. and {Lasenby}, A. and {Lavonen}, N. and {Lawrence}, C.~R. and {Leto}, C. and {Lindfors}, E. and {Mandolesi}, N. and {Massardi}, M. and {Max-Moerbeck}, W. and {Michelson}, P.~F. and {Mingaliev}, M. and {Natoli}, P. and {Nestoras}, I. and {Nieppola}, E. and {Nilsson}, K. and {Partridge}, B. and {Pavlidou}, V. and {Pearson}, T.~J. and {Procopio}, P. and {Rachen}, J.~P. and {Readhead}, A. and {Reeves}, R. and {Reimer}, A. and {Reinthal}, R. and {Ricciardi}, S. and {Richards}, J. and {Riquelme}, D. and {Saarinen}, J. and {Sajina}, A. and {Sandri}, M. and {Savolainen}, P. and {Sievers}, A. and {Sillanp{\"a}{\"a}}, A. and {Sotnikova}, Y. and {Stevenson}, M. and {Tagliaferri}, G. and {Takalo}, L. and {Tammi}, J. and {Tavagnacco}, D. and {Terenzi}, L. and {Toffolatti}, L. and {Tornikoski}, M. and {Trigilio}, C. and {Turunen}, M. and {Umana}, G. and {Ungerechts}, H. and {Villa}, F. and {Wu}, J. and {Zacchei}, A. and {Zensus}, J.~A. and {Zhou}, X.},
        title = "{Simultaneous Planck, Swift, and Fermi observations of X-ray and {\ensuremath{\gamma}}-ray selected blazars}",
      journal = {\aap},
         year = 2012,
        month = may,
       volume = {541},
          eid = {A160},
        pages = {A160},
          doi = {10.1051/0004-6361/201117825},
archivePrefix = {arXiv},
       eprint = {1108.1114},
 primaryClass = {astro-ph.CO},
       adsurl = {https://ui.adsabs.harvard.edu/abs/2012A&A...541A.160G}
}

@ARTICLE{Giommi1995A&AS,
       author = {{Giommi}, P. and {Ansari}, S.~G. and {Micol}, A.},
        title = "{Radio to X-ray energy distribution of BL Lacertae objects.}",
      journal = {\aaps},
         year = 1995,
        month = feb,
       volume = {109},
        pages = {267-291},
       adsurl = {https://ui.adsabs.harvard.edu/abs/1995A&AS..109..267G}
}

@ARTICLE{Wu2007A&A,
       author = {{Wu}, Zhongzu and {Jiang}, D.~R. and {Gu}, Minfeng and {Liu}, Yi},
        title = "{VLBI observations of seven BL Lacertae objects from RGB sample}",
      journal = {\aap},
         year = 2007,
        month = apr,
       volume = {466},
       number = {1},
        pages = {63-73},
          doi = {10.1051/0004-6361:20066754},
archivePrefix = {arXiv},
       eprint = {0706.0191},
 primaryClass = {astro-ph},
       adsurl = {https://ui.adsabs.harvard.edu/abs/2007A&A...466...63W}
}

@ARTICLE{Padovani1995ApJ,
       author = {{Padovani}, Paolo and {Giommi}, Paolo},
        title = "{The Connection between X-Ray-- and Radio-selected BL Lacertae Objects}",
      journal = {\apj},
         year = 1995,
        month = may,
       volume = {444},
        pages = {567},
          doi = {10.1086/175631},
archivePrefix = {arXiv},
       eprint = {astro-ph/9412073},
 primaryClass = {astro-ph},
       adsurl = {https://ui.adsabs.harvard.edu/abs/1995ApJ...444..567P}
}

@ARTICLE{Bai2001ApJ,
       author = {{Bai}, J.~M. and {Lee}, Myung Gyoon},
        title = "{New Evidence for the Unified Scheme of BL Lacertae Objects and FR I Radio Galaxies}",
      journal = {\apj},
         year = 2001,
        month = feb,
       volume = {548},
       number = {1},
        pages = {244-248},
          doi = {10.1086/318695},
archivePrefix = {arXiv},
       eprint = {astro-ph/0012340},
 primaryClass = {astro-ph},
       adsurl = {https://ui.adsabs.harvard.edu/abs/2001ApJ...548..244B}
}

@ARTICLE{Ledden1985ApJ,
       author = {{Ledden}, J.~E. and {O'Dell}, S.~L.},
        title = "{The radio-optical-X-ray spectral flux distributions of blazars.}",
      journal = {\apj},
         year = 1985,
        month = nov,
       volume = {298},
        pages = {630-643},
          doi = {10.1086/163647},
       adsurl = {https://ui.adsabs.harvard.edu/abs/1985ApJ...298..630L}
}

@ARTICLE{Bottcher2002ApJ,
       author = {{B{\"o}ttcher}, M. and {Dermer}, C.~D.},
        title = "{An Evolutionary Scenario for Blazar Unification}",
      journal = {\apj},
         year = 2002,
        month = jan,
       volume = {564},
       number = {1},
        pages = {86-91},
          doi = {10.1086/324134},
archivePrefix = {arXiv},
       eprint = {astro-ph/0106395},
 primaryClass = {astro-ph},
       adsurl = {https://ui.adsabs.harvard.edu/abs/2002ApJ...564...86B}
}

@ARTICLE{Wang2002ApJ,
       author = {{Wang}, Jian-Min and {Staubert}, R{\"u}diger and {Ho}, Luis C.},
        title = "{The Accretion Rates and Spectral Energy Distributions of BL Lacertae Objects}",
      journal = {\apj},
         year = 2002,
        month = nov,
       volume = {579},
       number = {2},
        pages = {554-559},
          doi = {10.1086/342875},
archivePrefix = {arXiv},
       eprint = {astro-ph/0207305},
 primaryClass = {astro-ph},
       adsurl = {https://ui.adsabs.harvard.edu/abs/2002ApJ...579..554W}
}

@ARTICLE{Paliya2021ApJS,
       author = {{Paliya}, Vaidehi S. and {Dom{\'\i}nguez}, A. and {Ajello}, M. and {Olmo-Garc{\'\i}a}, A. and {Hartmann}, D.},
        title = "{The Central Engines of Fermi Blazars}",
      journal = {\apjs},
         year = 2021,
        month = apr,
       volume = {253},
       number = {2},
          eid = {46},
        pages = {46},
          doi = {10.3847/1538-4365/abe135},
archivePrefix = {arXiv},
       eprint = {2101.10849},
 primaryClass = {astro-ph.HE},
       adsurl = {https://ui.adsabs.harvard.edu/abs/2021ApJS..253...46P}
}

@ARTICLE{Keenan2021MN,
       author = {{Keenan}, Mary and {Meyer}, Eileen T. and {Georganopoulos}, Markos and {Reddy}, Karthik and {French}, Omar J.},
        title = "{The relativistic jet dichotomy and the end of the blazar sequence}",
      journal = {\mnras},
         year = 2021,
        month = aug,
       volume = {505},
       number = {4},
        pages = {4726-4745},
          doi = {10.1093/mnras/stab1182},
archivePrefix = {arXiv},
       eprint = {2007.12661},
 primaryClass = {astro-ph.GA},
       adsurl = {https://ui.adsabs.harvard.edu/abs/2021MNRAS.505.4726K}
}

@ARTICLE{Meyer2011ApJ,
       author = {{Meyer}, Eileen T. and {Fossati}, Giovanni and {Georganopoulos}, Markos and {Lister}, Matthew L.},
        title = "{From the Blazar Sequence to the Blazar Envelope: Revisiting the Relativistic Jet Dichotomy in Radio-loud Active Galactic Nuclei}",
      journal = {\apj},
         year = 2011,
        month = oct,
       volume = {740},
       number = {2},
          eid = {98},
        pages = {98},
          doi = {10.1088/0004-637X/740/2/98},
archivePrefix = {arXiv},
       eprint = {1107.5105},
 primaryClass = {astro-ph.CO},
       adsurl = {https://ui.adsabs.harvard.edu/abs/2011ApJ...740...98M}
}

@ARTICLE{Ye2025A&A,
       author = {{Ye}, Xu-Hong and {Baldi}, Ranieri D. and {Chen}, Yong-Yun and {Bastieri}, Denis and {Fan}, Jun-Hui},
        title = "{Accretion properties and jet mechanisms for the low-excitation radio galaxies}",
      journal = {\aap},
         year = 2025,
        month = may,
       volume = {697},
          eid = {A176},
        pages = {A176},
          doi = {10.1051/0004-6361/202453260},
archivePrefix = {arXiv},
       eprint = {2503.23607},
 primaryClass = {astro-ph.HE},
       adsurl = {https://ui.adsabs.harvard.edu/abs/2025A&A...697A.176Y}
}

@ARTICLE{Nieppola2008A&A,
       author = {{Nieppola}, E. and {Valtaoja}, E. and {Tornikoski}, M. and {Hovatta}, T. and {Kotiranta}, M.},
        title = "{Blazar sequence - an artefact of Doppler boosting}",
      journal = {\aap},
         year = 2008,
        month = sep,
       volume = {488},
       number = {3},
        pages = {867-872},
          doi = {10.1051/0004-6361:200809716},
archivePrefix = {arXiv},
       eprint = {0803.0654},
 primaryClass = {astro-ph},
       adsurl = {https://ui.adsabs.harvard.edu/abs/2008A&A...488..867N}
}

@ARTICLE{Yang2022ApJ,
       author = {{Yang}, W.~X. and {Wang}, H.~G. and {Liu}, Y. and {Yang}, J.~H. and {Xiao}, H.~B. and {Ye}, X.~H. and {Pei}, Z.~Y. and {Zhang}, L.~X. and {Fan}, J.~H.},
        title = "{Beaming Effect in Fermi Blazars}",
      journal = {\apj},
         year = 2022,
        month = feb,
       volume = {925},
       number = {2},
          eid = {120},
        pages = {120},
          doi = {10.3847/1538-4357/ac3a09},
       adsurl = {https://ui.adsabs.harvard.edu/abs/2022ApJ...925..120Y}
}

@ARTICLE{Long2025ApJ,
       author = {{Long}, Qing-Chen and {Dong}, Ai-Jun and {Zhi}, Qi-Jun and {Shang}, Lun-Hua},
        title = "{Revisiting the Fundamental Planes of Black Hole Activity for Strong Jet Sources}",
      journal = {\apj},
         year = 2025,
        month = feb,
       volume = {980},
       number = {2},
          eid = {187},
        pages = {187},
          doi = {10.3847/1538-4357/adaaee},
archivePrefix = {arXiv},
       eprint = {2501.11381},
 primaryClass = {astro-ph.HE},
       adsurl = {https://ui.adsabs.harvard.edu/abs/2025ApJ...980..187L}
}

@ARTICLE{Jorstad2001ApJS,
       author = {{Jorstad}, Svetlana G. and {Marscher}, Alan P. and {Mattox}, John R. and {Wehrle}, Ann E. and {Bloom}, Steven D. and {Yurchenko}, Alexei V.},
        title = "{Multiepoch Very Long Baseline Array Observations of EGRET-detected Quasars and BL Lacertae Objects: Superluminal Motion of Gamma-Ray Bright Blazars}",
      journal = {\apjs},
         year = 2001,
        month = jun,
       volume = {134},
       number = {2},
        pages = {181-240},
          doi = {10.1086/320858},
archivePrefix = {arXiv},
       eprint = {astro-ph/0101570},
 primaryClass = {astro-ph},
       adsurl = {https://ui.adsabs.harvard.edu/abs/2001ApJS..134..181J}
}

@ARTICLE{Georganopoulos2003ApJ,
       author = {{Georganopoulos}, Markos and {Kazanas}, Demosthenes},
        title = "{Decelerating Flows in TeV Blazars: A Resolution to the BL Lacertae-FR I Unification Problem}",
      journal = {\apjl},
         year = 2003,
        month = sep,
       volume = {594},
       number = {1},
        pages = {L27-L30},
          doi = {10.1086/378557},
archivePrefix = {arXiv},
       eprint = {astro-ph/0307404},
 primaryClass = {astro-ph},
       adsurl = {https://ui.adsabs.harvard.edu/abs/2003ApJ...594L..27G}
}

@ARTICLE{Ghisellini2005A&A,
       author = {{Ghisellini}, G. and {Tavecchio}, F. and {Chiaberge}, M.},
        title = "{Structured jets in TeV BL Lac objects and radiogalaxies.  Implications for the observed properties}",
      journal = {\aap},
         year = 2005,
        month = mar,
       volume = {432},
       number = {2},
        pages = {401-410},
          doi = {10.1051/0004-6361:20041404},
archivePrefix = {arXiv},
       eprint = {astro-ph/0406093},
 primaryClass = {astro-ph},
       adsurl = {https://ui.adsabs.harvard.edu/abs/2005A&A...432..401G}
}

@ARTICLE{Piner2010ApJ,
       author = {{Piner}, B. Glenn and {Pant}, Niraj and {Edwards}, Philip G.},
        title = "{The Jets of TeV Blazars at Higher Resolution: 43 GHz and Polarimetric VLBA Observations from 2005 to 2009}",
      journal = {\apj},
         year = 2010,
        month = nov,
       volume = {723},
       number = {2},
        pages = {1150-1167},
          doi = {10.1088/0004-637X/723/2/1150},
archivePrefix = {arXiv},
       eprint = {1009.2269},
 primaryClass = {astro-ph.CO},
       adsurl = {https://ui.adsabs.harvard.edu/abs/2010ApJ...723.1150P}
}

@ARTICLE{Karamanavis2016A&A,
       author = {{Karamanavis}, V. and {Fuhrmann}, L. and {Krichbaum}, T.~P. and {Angelakis}, E. and {Hodgson}, J. and {Nestoras}, I. and {Myserlis}, I. and {Zensus}, J.~A. and {Sievers}, A. and {Ciprini}, S.},
        title = "{PKS 1502+106: A high-redshift Fermi blazar at extreme angular resolution. Structural dynamics with VLBI imaging up to 86 GHz}",
      journal = {\aap},
         year = 2016,
        month = feb,
       volume = {586},
          eid = {A60},
        pages = {A60},
          doi = {10.1051/0004-6361/201527225},
archivePrefix = {arXiv},
       eprint = {1511.01085},
 primaryClass = {astro-ph.HE},
       adsurl = {https://ui.adsabs.harvard.edu/abs/2016A&A...586A..60K}
}

@ARTICLE{Balmaverde2006A&A,
       author = {{Balmaverde}, B. and {Capetti}, A. and {Grandi}, P.},
        title = "{The Chandra view of the 3C/FR I sample of low luminosity radio-galaxies}",
      journal = {\aap},
         year = 2006,
        month = may,
       volume = {451},
       number = {1},
        pages = {35-44},
          doi = {10.1051/0004-6361:20053799},
archivePrefix = {arXiv},
       eprint = {astro-ph/0601175},
 primaryClass = {astro-ph},
       adsurl = {https://ui.adsabs.harvard.edu/abs/2006A&A...451...35B}
}

@ARTICLE{Shen2014Nat,
       author = {{Shen}, Yue and {Ho}, Luis C.},
        title = "{The diversity of quasars unified by accretion and orientation}",
      journal = {\nat},
         year = 2014,
        month = sep,
       volume = {513},
       number = {7517},
        pages = {210-213},
          doi = {10.1038/nature13712},
archivePrefix = {arXiv},
       eprint = {1409.2887},
 primaryClass = {astro-ph.GA},
       adsurl = {https://ui.adsabs.harvard.edu/abs/2014Natur.513..210S}
}

@ARTICLE{Chiaberge2000A&A,
       author = {{Chiaberge}, M. and {Celotti}, A. and {Capetti}, A. and {Ghisellini}, G.},
        title = "{Does the unification of BL Lac and FR I radio galaxies require jet velocity structures?}",
      journal = {\aap},
         year = 2000,
        month = jun,
       volume = {358},
        pages = {104-112},
          doi = {10.48550/arXiv.astro-ph/0003197},
archivePrefix = {arXiv},
       eprint = {astro-ph/0003197},
 primaryClass = {astro-ph},
       adsurl = {https://ui.adsabs.harvard.edu/abs/2000A&A...358..104C}
}

@ARTICLE{Capetti2000MN,
       author = {{Capetti}, A. and {Trussoni}, E. and {Celotti}, A. and {Feretti}, L. and {Chiaberge}, M.},
        title = "{Spectral energy distributions of FR I nuclei and the FR I/BL Lac unifying model}",
      journal = {\mnras},
         year = 2000,
        month = oct,
       volume = {318},
       number = {2},
        pages = {493-500},
          doi = {10.1046/j.1365-8711.2000.03823.x},
       adsurl = {https://ui.adsabs.harvard.edu/abs/2000MNRAS.318..493C}
}

@ARTICLE{Trussoni2003A&A,
       author = {{Trussoni}, E. and {Capetti}, A. and {Celotti}, A. and {Chiaberge}, M. and {Feretti}, L.},
        title = "{A multi-wavelength test of the FR I-BL Lac unifying model}",
      journal = {\aap},
         year = 2003,
        month = jun,
       volume = {403},
        pages = {889-899},
          doi = {10.1051/0004-6361:20030417},
archivePrefix = {arXiv},
       eprint = {astro-ph/0304018},
 primaryClass = {astro-ph},
       adsurl = {https://ui.adsabs.harvard.edu/abs/2003A&A...403..889T}
}

@ARTICLE{Wright2006PASP,
       author = {{Wright}, E.~L.},
        title = "{A Cosmology Calculator for the World Wide Web}",
      journal = {\pasp},
         year = 2006,
        month = dec,
       volume = {118},
       number = {850},
        pages = {1711-1715},
          doi = {10.1086/510102},
archivePrefix = {arXiv},
       eprint = {astro-ph/0609593},
 primaryClass = {astro-ph},
       adsurl = {https://ui.adsabs.harvard.edu/abs/2006PASP..118.1711W}
}

@ARTICLE{Hovatta2009,
       author = {{Hovatta}, T. and {Valtaoja}, E. and {Tornikoski}, M. and {L{\"a}hteenm{\"a}ki}, A.},
        title = "{Doppler factors, Lorentz factors and viewing angles for quasars, BL Lacertae objects and radio galaxies}",
      journal = {\aap},
         year = 2009,
        month = feb,
       volume = {494},
       number = {2},
        pages = {527-537},
          doi = {10.1051/0004-6361:200811150},
archivePrefix = {arXiv},
       eprint = {0811.4278},
 primaryClass = {astro-ph},
       adsurl = {https://ui.adsabs.harvard.edu/abs/2009A&A...494..527H}
}

@ARTICLE{Wu2014,
       author = {{Wu}, Zhongzu and {Jiang}, Dongrong and {Gu}, Minfeng and {Chen}, Liang},
        title = "{Why are some BL Lacertaes detected by Fermi, but others not?}",
      journal = {\aap},
         year = 2014,
        month = feb,
       volume = {562},
          eid = {A64},
        pages = {A64},
          doi = {10.1051/0004-6361/201220851},
archivePrefix = {arXiv},
       eprint = {1401.0652},
 primaryClass = {astro-ph.HE},
       adsurl = {https://ui.adsabs.harvard.edu/abs/2014A&A...562A..64W}
}

@ARTICLE{Liodakis2017MN,
       author = {{Liodakis}, I. and {Marchili}, N. and {Angelakis}, E. and {Fuhrmann}, L. and {Nestoras}, I. and {Myserlis}, I. and {Karamanavis}, V. and {Krichbaum}, T.~P. and {Sievers}, A. and {Ungerechts}, H. and {Zensus}, J.~A.},
        title = "{F-GAMMA: variability Doppler factors of blazars from multiwavelength monitoring}",
      journal = {\mnras},
         year = 2017,
        month = apr,
       volume = {466},
       number = {4},
        pages = {4625-4632},
          doi = {10.1093/mnras/stx002},
archivePrefix = {arXiv},
       eprint = {1701.01452},
 primaryClass = {astro-ph.HE},
       adsurl = {https://ui.adsabs.harvard.edu/abs/2017MNRAS.466.4625L}
}

@ARTICLE{Liodakis2018ApJ,
       author = {{Liodakis}, Ioannis and {Hovatta}, Talvikki and {Huppenkothen}, Daniela and {Kiehlmann}, Sebastian and {Max-Moerbeck}, Walter and {Readhead}, Anthony C.~S.},
        title = "{Constraining the Limiting Brightness Temperature and Doppler Factors for the Largest Sample of Radio-bright Blazars}",
      journal = {\apj},
         year = 2018,
        month = oct,
       volume = {866},
       number = {2},
          eid = {137},
        pages = {137},
          doi = {10.3847/1538-4357/aae2b7},
archivePrefix = {arXiv},
       eprint = {1809.08249},
 primaryClass = {astro-ph.HE},
       adsurl = {https://ui.adsabs.harvard.edu/abs/2018ApJ...866..137L}
}

@ARTICLE{Ye2021PASJ,
       author = {{Ye}, Xu-Hong and {Fan}, Jun-Hui},
        title = "{Unification of BL Lac objects and FR I and FR II(G) radio galaxies, and Doppler factor estimation for BL Lac objects}",
      journal = {\pasj},
         year = 2021,
        month = aug,
       volume = {73},
       number = {4},
        pages = {775-785},
          doi = {10.1093/pasj/psab039},
archivePrefix = {arXiv},
       eprint = {2111.09023},
 primaryClass = {astro-ph.GA},
       adsurl = {https://ui.adsabs.harvard.edu/abs/2021PASJ...73..775Y}
}

@ARTICLE{Ajello2022ApJS,
       author = {{Ajello}, M. and {Baldini}, L. and {Ballet}, J. and {Bastieri}, D. and {Becerra Gonzalez}, J. and {Bellazzini}, R. and {Berretta}, A. and {Bissaldi}, E. and {Bonino}, R. and {Brill}, A. and {Bruel}, P. and {Buson}, S. and {Caputo}, R. and {Caraveo}, P.~A. and {Cheung}, C.~C. and {Chiaro}, G. and {Cibrario}, N. and {Ciprini}, S. and {Crnogorcevic}, M. and {Cutini}, S. and {D'Ammando}, F. and {De Gaetano}, S. and {Di Lalla}, N. and {Di Venere}, L. and {Dom{\'\i}nguez}, A. and {Ramazani}, V. Fallah and {Ferrara}, E.~C. and {Fiori}, A. and {Fukazawa}, Y. and {Funk}, S. and {Fusco}, P. and {Gammaldi}, V. and {Gargano}, F. and {Garrappa}, S. and {Gasparrini}, D. and {Giglietto}, N. and {Giordano}, F. and {Giroletti}, M. and {Green}, D. and {Grenier}, I.~A. and {Guiriec}, S. and {Horan}, D. and {Hou}, X. and {Kayanoki}, T. and {Kuss}, M. and {Larsson}, S. and {Latronico}, L. and {Lewis}, T. and {Li}, J. and {Liodakis}, I. and {Longo}, F. and {Loparco}, F. and {Lott}, B. and {Lovellette}, M.~N. and {Lubrano}, P. and {Madejski}, G.~M. and {Maldera}, S. and {Manfreda}, A. and {Mart{\'\i}-Devesa}, G. and {Mazziotta}, M.~N. and {Mereu}, I. and {Michelson}, P.~F. and {Mirabal}, N. and {Mitthumsiri}, W. and {Mizuno}, T. and {Monzani}, M.~E. and {Morselli}, A. and {Moskalenko}, I.~V. and {Negro}, M. and {Ojha}, R. and {Orienti}, M. and {Orlando}, E. and {Ormes}, J.~F. and {Pei}, Z. and {Pe{\~n}a-Herazo}, H. and {Persic}, M. and {Pesce-Rollins}, M. and {Petrosian}, V. and {Pillera}, R. and {Poon}, H. and {Porter}, T.~A. and {Principe}, G. and {Rain{\`o}}, S. and {Rando}, R. and {Rani}, B. and {Razzano}, M. and {Razzaque}, S. and {Reimer}, A. and {Reimer}, O. and {Scotton}, L. and {Serini}, D. and {Sgr{\`o}}, C. and {Siskind}, E.~J. and {Spandre}, G. and {Spinelli}, P. and {Suson}, D.~J. and {Tajima}, H. and {Torres}, D.~F. and {Valverde}, J. and {Yassin}, H. and {Zaharijas}, G.},
        title = "{The Fourth Catalog of Active Galactic Nuclei Detected by the Fermi Large Area Telescope: Data Release 3}",
      journal = {\apjs},
         year = 2022,
        month = dec,
       volume = {263},
       number = {2},
          eid = {24},
        pages = {24},
          doi = {10.3847/1538-4365/ac9523},
archivePrefix = {arXiv},
       eprint = {2209.12070},
 primaryClass = {astro-ph.HE},
       adsurl = {https://ui.adsabs.harvard.edu/abs/2022ApJS..263...24A}
}

@ARTICLE{Xiong2015MN-a,
       author = {{Xiong}, Dingrong and {Zhang}, Xiong and {Bai}, Jinming and {Zhang}, Haojing},
        title = "{Basic properties of Fermi blazars and the `blazar sequence'}",
      journal = {\mnras},
         year = 2015,
        month = jul,
       volume = {450},
       number = {4},
        pages = {3568-3578},
          doi = {10.1093/mnras/stv812},
archivePrefix = {arXiv},
       eprint = {1504.02706},
 primaryClass = {astro-ph.HE},
       adsurl = {https://ui.adsabs.harvard.edu/abs/2015MNRAS.450.3568X}
}

@ARTICLE{Xiong2015MN-b,
       author = {{Xiong}, Dingrong and {Zhang}, Xiong and {Bai}, Jinming and {Zhang}, Haojing},
        title = "{From the `blazar sequence' to unification of blazars and radio galaxies}",
      journal = {\mnras},
         year = 2015,
        month = aug,
       volume = {451},
       number = {3},
        pages = {2750-2756},
          doi = {10.1093/mnras/stv1038},
archivePrefix = {arXiv},
       eprint = {1505.01408},
 primaryClass = {astro-ph.HE},
       adsurl = {https://ui.adsabs.harvard.edu/abs/2015MNRAS.451.2750X}
}

@ARTICLE{Chang2019A&A,
       author = {{Chang}, Y.-L. and {Arsioli}, B. and {Giommi}, P. and {Padovani}, P. and {Brandt}, C.~H.},
        title = "{The 3HSP catalogue of extreme and high-synchrotron peaked blazars}",
      journal = {\aap},
         year = 2019,
        month = dec,
       volume = {632},
          eid = {A77},
        pages = {A77},
          doi = {10.1051/0004-6361/201834526},
archivePrefix = {arXiv},
       eprint = {1909.08279},
 primaryClass = {astro-ph.HE},
       adsurl = {https://ui.adsabs.harvard.edu/abs/2019A&A...632A..77C}
}

@ARTICLE{Donato2005A&A,
       author = {{Donato}, D. and {Sambruna}, R.~M. and {Gliozzi}, M.},
        title = "{Six years of BeppoSAX observations of blazars: A spectral catalog}",
      journal = {\aap},
         year = 2005,
        month = apr,
       volume = {433},
       number = {3},
        pages = {1163-1169},
          doi = {10.1051/0004-6361:20034555},
archivePrefix = {arXiv},
       eprint = {physics/0412114},
 primaryClass = {physics.data-an},
       adsurl = {https://ui.adsabs.harvard.edu/abs/2005A&A...433.1163D}
}

@ARTICLE{Panessa2019NatAs,
       author = {{Panessa}, Francesca and {Baldi}, Ranieri Diego and {Laor}, Ari and {Padovani}, Paolo and {Behar}, Ehud and {McHardy}, Ian},
        title = "{The origin of radio emission from radio-quiet active galactic nuclei}",
      journal = {Nature Astronomy},
         year = 2019,
        month = apr,
       volume = {3},
        pages = {387-396},
          doi = {10.1038/s41550-019-0765-4},
archivePrefix = {arXiv},
       eprint = {1902.05917},
 primaryClass = {astro-ph.GA},
       adsurl = {https://ui.adsabs.harvard.edu/abs/2019NatAs...3..387P}
}

@ARTICLE{Cavagnolo2010ApJ,
       author = {{Cavagnolo}, K.~W. and {McNamara}, B.~R. and {Nulsen}, P.~E.~J. and {Carilli}, C.~L. and {Jones}, C. and {B{\^\i}rzan}, L.},
        title = "{A Relationship Between AGN Jet Power and Radio Power}",
      journal = {\apj},
         year = 2010,
        month = sep,
       volume = {720},
       number = {2},
        pages = {1066-1072},
          doi = {10.1088/0004-637X/720/2/1066},
archivePrefix = {arXiv},
       eprint = {1006.5699},
 primaryClass = {astro-ph.CO},
       adsurl = {https://ui.adsabs.harvard.edu/abs/2010ApJ...720.1066C}
}

@ARTICLE{Ghisellini1993ApJ,
       author = {{Ghisellini}, G. and {Padovani}, P. and {Celotti}, A. and {Maraschi}, L.},
        title = "{Relativistic Bulk Motion in Active Galactic Nuclei}",
      journal = {\apj},
         year = 1993,
        month = apr,
       volume = {407},
        pages = {65},
          doi = {10.1086/172493},
       adsurl = {https://ui.adsabs.harvard.edu/abs/1993ApJ...407...65G}
}

@ARTICLE{Fan2014RAA,
       author = {{Fan}, Jun-Hui and {Bastieri}, Denis and {Yang}, Jiang-He and {Liu}, Yi and {Hua}, Tong-Xu and {Yuan}, Yu-Hai and {Wu}, De-Xiang},
        title = "{The lower limit of the Doppler factor for a Fermi blazar sample}",
      journal = {RAA},
         year = 2014,
        month = sep,
       volume = {14},
       number = {9},
          eid = {1135-1145},
        pages = {1135-1145},
          doi = {10.1088/1674-4527/14/9/004},
       adsurl = {https://ui.adsabs.harvard.edu/abs/2014RAA....14.1135F}
}

@ARTICLE{Pei2022ApJ,
       author = {{Pei}, Zhiyuan and {Fan}, Junhui and {Yang}, Jianghe and {Huang}, Danyi and {Li}, Ziyan},
        title = "{The Estimation of Fundamental Physics Parameters for Fermi-LAT Blazars}",
      journal = {\apj},
         year = 2022,
        month = jan,
       volume = {925},
       number = {1},
          eid = {97},
        pages = {97},
          doi = {10.3847/1538-4357/ac3aeb},
archivePrefix = {arXiv},
       eprint = {2112.00530},
 primaryClass = {astro-ph.HE},
       adsurl = {https://ui.adsabs.harvard.edu/abs/2022ApJ...925...97P}
}

@ARTICLE{Fan1994Ap&SS,
       author = {{Fan}, J.~H. and {Huang}, Z.~H. and {Li}, J.~J. and {Xie}, G.~Z. and {Zhang}, J.~Y.},
        title = "{BL Lac Objects and Acceleration Model}",
      journal = {\apss},
         year = 1994,
        month = mar,
       volume = {213},
       number = {2},
        pages = {305-316},
          doi = {10.1007/BF00658217},
       adsurl = {https://ui.adsabs.harvard.edu/abs/1994Ap&SS.213..305F}
}

@ARTICLE{Long2026MN,
       author = {{Long}, Qing-Chen and {Dong}, Ai-Jun and {Zhi}, Qi-Jun},
        title = "{The fundamental planes of black hole activity for high-synchrotron-peaked BL Lacertae objects}",
      journal = {\mnras},
         year = 2026,
        month = may,
       volume = {548},
       number = {1},
          eid = {stag608},
        pages = {stag608},
          doi = {10.1093/mnras/stag608},
archivePrefix = {arXiv},
       eprint = {2606.21426},
 primaryClass = {astro-ph.HE},
       adsurl = {https://ui.adsabs.harvard.edu/abs/2026MNRAS.548ag608L}
}

@ARTICLE{Ghisellini1998MN,
       author = {{Ghisellini}, G. and {Celotti}, A. and {Fossati}, G. and {Maraschi}, L. and {Comastri}, A.},
        title = "{A theoretical unifying scheme for gamma-ray bright blazars}",
      journal = {\mnras},
         year = 1998,
        month = dec,
       volume = {301},
       number = {2},
        pages = {451-468},
          doi = {10.1046/j.1365-8711.1998.02032.x},
archivePrefix = {arXiv},
       eprint = {astro-ph/9807317},
 primaryClass = {astro-ph},
       adsurl = {https://ui.adsabs.harvard.edu/abs/1998MNRAS.301..451G}
}

@ARTICLE{Ghisellini2014Nat,
       author = {{Ghisellini}, G. and {Tavecchio}, F. and {Maraschi}, L. and {Celotti}, A. and {Sbarrato}, T.},
        title = "{The power of relativistic jets is larger than the luminosity of their accretion disks}",
      journal = {\nat},
         year = 2014,
        month = nov,
       volume = {515},
       number = {7527},
        pages = {376-378},
          doi = {10.1038/nature13856},
archivePrefix = {arXiv},
       eprint = {1411.5368},
 primaryClass = {astro-ph.HE},
       adsurl = {https://ui.adsabs.harvard.edu/abs/2014Natur.515..376G}
}

@ARTICLE{Vestergaard2006ApJ,
       author = {{Vestergaard}, Marianne and {Peterson}, Bradley M.},
        title = "{Determining Central Black Hole Masses in Distant Active Galaxies and Quasars. II. Improved Optical and UV Scaling Relationships}",
      journal = {\apj},
         year = 2006,
        month = apr,
       volume = {641},
       number = {2},
        pages = {689-709},
          doi = {10.1086/500572},
archivePrefix = {arXiv},
       eprint = {astro-ph/0601303},
 primaryClass = {astro-ph},
       adsurl = {https://ui.adsabs.harvard.edu/abs/2006ApJ...641..689V}
}

@ARTICLE{Shaw2012ApJ,
       author = {{Shaw}, Michael S. and {Romani}, Roger W. and {Cotter}, Garret and {Healey}, Stephen E. and {Michelson}, Peter F. and {Readhead}, Anthony C.~S. and {Richards}, Joseph L. and {Max-Moerbeck}, Walter and {King}, Oliver G. and {Potter}, William J.},
        title = "{Spectroscopy of Broad-line Blazars from 1LAC}",
      journal = {\apj},
         year = 2012,
        month = mar,
       volume = {748},
       number = {1},
          eid = {49},
        pages = {49},
          doi = {10.1088/0004-637X/748/1/49},
archivePrefix = {arXiv},
       eprint = {1201.0999},
 primaryClass = {astro-ph.HE},
       adsurl = {https://ui.adsabs.harvard.edu/abs/2012ApJ...748...49S}
}

@ARTICLE{Gultekin2009ApJ,
       author = {{G{\"u}ltekin}, Kayhan and {Cackett}, Edward M. and {Miller}, Jon M. and {Di Matteo}, Tiziana and {Markoff}, Sera and {Richstone}, Douglas O.},
        title = "{The Fundamental Plane of Accretion onto Black Holes with Dynamical Masses}",
      journal = {\apj},
         year = 2009,
        month = nov,
       volume = {706},
       number = {1},
        pages = {404-416},
          doi = {10.1088/0004-637X/706/1/404},
archivePrefix = {arXiv},
       eprint = {0906.3285},
 primaryClass = {astro-ph.HE},
       adsurl = {https://ui.adsabs.harvard.edu/abs/2009ApJ...706..404G}
}

@ARTICLE{Bariuan2022MN,
       author = {{Bariuan}, Luis Gabriel C. and {Snios}, Bradford and {Sobolewska}, Ma{\l}gosia and {Siemiginowska}, Aneta and {Schwartz}, Daniel A.},
        title = "{The Fundamental Planes of black hole activity for radio-loud and radio-quiet quasars}",
      journal = {\mnras},
         year = 2022,
        month = jul,
       volume = {513},
       number = {4},
        pages = {4673-4681},
          doi = {10.1093/mnras/stac1153},
archivePrefix = {arXiv},
       eprint = {2201.04666},
 primaryClass = {astro-ph.HE},
       adsurl = {https://ui.adsabs.harvard.edu/abs/2022MNRAS.513.4673B}
}

@ARTICLE{Chen2024ApJS,
       author = {{Chen}, Guohai and {Zheng}, Zepeng and {Zeng}, Xiangtao and {Zhang}, Lixia and {Xiao}, Hubing and {Liu}, Xiang and {Cui}, Lang and {Fan}, Junhui},
        title = "{A Study of Broad Emission Line and Doppler Factor Estimation for Fermi Blazars}",
      journal = {\apjs},
         year = 2024,
        month = mar,
       volume = {271},
       number = {1},
          eid = {20},
        pages = {20},
          doi = {10.3847/1538-4365/ad1c67},
       adsurl = {https://ui.adsabs.harvard.edu/abs/2024ApJS..271...20C}
}

@ARTICLE{Sikora2007ApJ,
       author = {{Sikora}, Marek and {Stawarz}, {\L}ukasz and {Lasota}, Jean-Pierre},
        title = "{Radio Loudness of Active Galactic Nuclei: Observational Facts and Theoretical Implications}",
      journal = {\apj},
         year = 2007,
        month = apr,
       volume = {658},
       number = {2},
        pages = {815-828},
          doi = {10.1086/511972},
archivePrefix = {arXiv},
       eprint = {astro-ph/0604095},
 primaryClass = {astro-ph},
       adsurl = {https://ui.adsabs.harvard.edu/abs/2007ApJ...658..815S}
}

@ARTICLE{Chai2012ApJ,
       author = {{Chai}, Bo and {Cao}, Xinwu and {Gu}, Minfeng},
        title = "{What Governs the Bulk Velocity of the Jet Components in Active Galactic Nuclei?}",
      journal = {\apj},
         year = 2012,
        month = nov,
       volume = {759},
       number = {2},
          eid = {114},
        pages = {114},
          doi = {10.1088/0004-637X/759/2/114},
archivePrefix = {arXiv},
       eprint = {1209.4702},
 primaryClass = {astro-ph.CO},
       adsurl = {https://ui.adsabs.harvard.edu/abs/2012ApJ...759..114C}
}

@ARTICLE{Zhou2009RAA,
       author = {{Zhou}, Ming and {Cao}, Xin-Wu},
        title = "{The relation between black hole masses and Lorentz factors of the jet components in blazars}",
      journal = {RAA},
         year = 2009,
        month = mar,
       volume = {9},
       number = {3},
        pages = {293-301},
          doi = {10.1088/1674-4527/9/3/003},
archivePrefix = {arXiv},
       eprint = {0806.2435},
 primaryClass = {astro-ph},
       adsurl = {https://ui.adsabs.harvard.edu/abs/2009RAA.....9..293Z}
}

@ARTICLE{Saxton2008,
       author = {{Saxton}, R.~D. and {Read}, A.~M. and {Esquej}, P. and {Freyberg}, M.~J. and {Altieri}, B. and {Bermejo}, D.},
        title = "{The first XMM-Newton slew survey catalogue: XMMSL1}",
      journal = {\aap},
         year = 2008,
        month = mar,
       volume = {480},
       number = {2},
        pages = {611-622},
          doi = {10.1051/0004-6361:20079193},
archivePrefix = {arXiv},
       eprint = {0801.3732},
 primaryClass = {astro-ph},
       adsurl = {https://ui.adsabs.harvard.edu/abs/2008A&A...480..611S}
}

@ARTICLE{Liu2006ApJ,
       author = {{Liu}, Yi and {Jiang}, Dong Rong and {Gu}, Min Feng},
        title = "{The Jet Power, Radio Loudness, and Black Hole Mass in Radio-loud Active Galactic Nuclei}",
      journal = {\apj},
         year = 2006,
        month = feb,
       volume = {637},
       number = {2},
        pages = {669-681},
          doi = {10.1086/498639},
archivePrefix = {arXiv},
       eprint = {astro-ph/0510241},
 primaryClass = {astro-ph},
       adsurl = {https://ui.adsabs.harvard.edu/abs/2006ApJ...637..669L}
}

@ARTICLE{Wang2006ApJ,
       author = {{Wang}, Ran and {Wu}, Xue-Bing and {Kong}, Min-Zhi},
        title = "{The Black Hole Fundamental Plane from a Uniform Sample of Radio and X-Ray-emitting Broad-Line AGNs}",
      journal = {\apj},
         year = 2006,
        month = jul,
       volume = {645},
       number = {2},
        pages = {890-899},
          doi = {10.1086/504401},
archivePrefix = {arXiv},
       eprint = {astro-ph/0603514},
 primaryClass = {astro-ph},
       adsurl = {https://ui.adsabs.harvard.edu/abs/2006ApJ...645..890W}
}

@ARTICLE{Dwelly2017MN,
       author = {{Dwelly}, T. and {Salvato}, M. and {Merloni}, A. and {Brusa}, M. and {Buchner}, J. and {Anderson}, S.~F. and {Boller}, Th. and {Brandt}, W.~N. and {Budav{\'a}ri}, T. and {Clerc}, N. and {Coffey}, D. and {Del Moro}, A. and {Georgakakis}, A. and {Green}, P.~J. and {Jin}, C. and {Menzel}, M.-L. and {Myers}, A.~D. and {Nandra}, K. and {Nichol}, R.~C. and {Ridl}, J. and {Schwope}, A.~D. and {Simm}, T.},
        title = "{SPIDERS: selection of spectroscopic targets using AGN candidates detected in all-sky X-ray surveys}",
      journal = {\mnras},
         year = 2017,
        month = jul,
       volume = {469},
       number = {1},
        pages = {1065-1095},
          doi = {10.1093/mnras/stx864},
archivePrefix = {arXiv},
       eprint = {1704.01796},
 primaryClass = {astro-ph.GA},
       adsurl = {https://ui.adsabs.harvard.edu/abs/2017MNRAS.469.1065D}
}

@ARTICLE{Shen2011ApJS,
       author = {{Shen}, Yue and {Richards}, Gordon T. and {Strauss}, Michael A. and {Hall}, Patrick B. and {Schneider}, Donald P. and {Snedden}, Stephanie and {Bizyaev}, Dmitry and {Brewington}, Howard and {Malanushenko}, Viktor and {Malanushenko}, Elena and {Oravetz}, Dan and {Pan}, Kaike and {Simmons}, Audrey},
        title = "{A Catalog of Quasar Properties from Sloan Digital Sky Survey Data Release 7}",
      journal = {\apjs},
         year = 2011,
        month = jun,
       volume = {194},
       number = {2},
          eid = {45},
        pages = {45},
          doi = {10.1088/0067-0049/194/2/45},
archivePrefix = {arXiv},
       eprint = {1006.5178},
 primaryClass = {astro-ph.CO},
       adsurl = {https://ui.adsabs.harvard.edu/abs/2011ApJS..194...45S}
}

@ARTICLE{Kurinsky2013,
       author = {{Kurinsky}, N. and {Sajina}, A. and {Partridge}, B. and {Myers}, S. and {Chen}, X. and {L{\'o}pez-Caniego}, M.},
        title = "{VLA/JVLA monitoring of bright northern radio sources}",
      journal = {\aap},
         year = 2013,
        month = jan,
       volume = {549},
          eid = {A133},
        pages = {A133},
          doi = {10.1051/0004-6361/201219851},
archivePrefix = {arXiv},
       eprint = {1211.3931},
 primaryClass = {astro-ph.CO},
       adsurl = {https://ui.adsabs.harvard.edu/abs/2013A&A...549A.133K}
}

@ARTICLE{Yuan2012ApJ,
       author = {{Yuan}, Zunli and {Wang}, Jiancheng},
        title = "{On the Evolution of the Cores of Radio Sources and Their Extended Radio Emission}",
      journal = {\apj},
         year = 2012,
        month = jan,
       volume = {744},
       number = {2},
          eid = {84},
        pages = {84},
          doi = {10.1088/0004-637X/744/2/84},
archivePrefix = {arXiv},
       eprint = {1109.4028},
 primaryClass = {astro-ph.CO},
       adsurl = {https://ui.adsabs.harvard.edu/abs/2012ApJ...744...84Y}
}

@ARTICLE{Gu2009MN,
       author = {{Gu}, Minfeng and {Cao}, Xinwu and {Jiang}, D.~R.},
        title = "{The bulk kinetic power of radio jets in active galactic nuclei}",
      journal = {\mnras},
         year = 2009,
        month = jun,
       volume = {396},
       number = {2},
        pages = {984-996},
          doi = {10.1111/j.1365-2966.2009.14758.x},
archivePrefix = {arXiv},
       eprint = {0903.1896},
 primaryClass = {astro-ph.GA},
       adsurl = {https://ui.adsabs.harvard.edu/abs/2009MNRAS.396..984G}
}

@ARTICLE{Mantovani2015,
       author = {{Mantovani}, F. and {Bondi}, M. and {Mack}, K.-H. and {Alef}, W. and {Ros}, E. and {Zensus}, J.~A.},
        title = "{A sample of weak blazars at milli-arcsecond resolution}",
      journal = {\aap},
         year = 2015,
        month = may,
       volume = {577},
          eid = {A36},
        pages = {A36},
          doi = {10.1051/0004-6361/201425527},
archivePrefix = {arXiv},
       eprint = {1502.07176},
 primaryClass = {astro-ph.GA},
       adsurl = {https://ui.adsabs.harvard.edu/abs/2015A&A...577A..36M}
}

@ARTICLE{Marshall2018ApJ,
       author = {{Marshall}, H.~L. and {Gelbord}, J.~M. and {Worrall}, D.~M. and {Birkinshaw}, M. and {Schwartz}, D.~A. and {Jauncey}, D.~L. and {Griffiths}, G. and {Murphy}, D.~W. and {Lovell}, J.~E.~J. and {Perlman}, E.~S. and {Godfrey}, L.},
        title = "{An X-Ray Imaging Survey of Quasar Jets: The Complete Survey}",
      journal = {\apj},
         year = 2018,
        month = mar,
       volume = {856},
       number = {1},
          eid = {66},
        pages = {66},
          doi = {10.3847/1538-4357/aaaf66},
archivePrefix = {arXiv},
       eprint = {1802.04714},
 primaryClass = {astro-ph.HE},
       adsurl = {https://ui.adsabs.harvard.edu/abs/2018ApJ...856...66M}
}

@ARTICLE{Sbarrato2012MN,
       author = {{Sbarrato}, T. and {Ghisellini}, G. and {Maraschi}, L. and {Colpi}, M.},
        title = "{The relation between broad lines and {\ensuremath{\gamma}}-ray luminosities in Fermi blazars}",
      journal = {\mnras},
         year = 2012,
        month = apr,
       volume = {421},
       number = {2},
        pages = {1764-1778},
          doi = {10.1111/j.1365-2966.2012.20442.x},
archivePrefix = {arXiv},
       eprint = {1108.0927},
 primaryClass = {astro-ph.HE},
       adsurl = {https://ui.adsabs.harvard.edu/abs/2012MNRAS.421.1764S}
}

@ARTICLE{Yuan2018ApJS,
       author = {{Yuan}, Zunli and {Wang}, Jiancheng and {Worrall}, D.~M. and {Zhang}, Bin-Bin and {Mao}, Jirong},
        title = "{Determining the Core Radio Luminosity Function of Radio AGNs via Copula}",
      journal = {\apjs},
         year = 2018,
        month = dec,
       volume = {239},
       number = {2},
          eid = {33},
        pages = {33},
          doi = {10.3847/1538-4365/aaed3b},
archivePrefix = {arXiv},
       eprint = {1810.12713},
 primaryClass = {astro-ph.GA},
       adsurl = {https://ui.adsabs.harvard.edu/abs/2018ApJS..239...33Y}
}

@ARTICLE{Kovalev2020MN,
       author = {{Kovalev}, Y.~Y. and {Pushkarev}, A.~B. and {Nokhrina}, E.~E. and {Plavin}, A.~V. and {Beskin}, V.~S. and {Chernoglazov}, A.~V. and {Lister}, M.~L. and {Savolainen}, T.},
        title = "{A transition from parabolic to conical shape as a common effect in nearby AGN jets}",
      journal = {\mnras},
         year = 2020,
        month = jul,
       volume = {495},
       number = {4},
        pages = {3576-3591},
          doi = {10.1093/mnras/staa1121},
archivePrefix = {arXiv},
       eprint = {1907.01485},
 primaryClass = {astro-ph.GA},
       adsurl = {https://ui.adsabs.harvard.edu/abs/2020MNRAS.495.3576K}
}

@ARTICLE{Woo2002ApJ,
       author = {{Woo}, Jong-Hak and {Urry}, C. Megan},
        title = "{Active Galactic Nucleus Black Hole Masses and Bolometric Luminosities}",
      journal = {\apj},
         year = 2002,
        month = nov,
       volume = {579},
       number = {2},
        pages = {530-544},
          doi = {10.1086/342878},
archivePrefix = {arXiv},
       eprint = {astro-ph/0207249},
 primaryClass = {astro-ph},
       adsurl = {https://ui.adsabs.harvard.edu/abs/2002ApJ...579..530W}
}

@ARTICLE{Pian2005MN,
       author = {{Pian}, E. and {Falomo}, R. and {Treves}, A.},
        title = "{Hubble Space Telescope ultraviolet spectroscopy of blazars: emission-line properties and black hole masses}",
      journal = {\mnras},
         year = 2005,
        month = aug,
       volume = {361},
       number = {3},
        pages = {919-926},
          doi = {10.1111/j.1365-2966.2005.09216.x},
archivePrefix = {arXiv},
       eprint = {astro-ph/0506725},
 primaryClass = {astro-ph},
       adsurl = {https://ui.adsabs.harvard.edu/abs/2005MNRAS.361..919P}
}

@ARTICLE{Nilsson2003,
       author = {{Nilsson}, K. and {Pursimo}, T. and {Heidt}, J. and {Takalo}, L.~O. and {Sillanp{\"a}{\"a}}, A. and {Brinkmann}, W.},
        title = "{R-band imaging of the host galaxies of RGB BL Lacertae objects}",
      journal = {\aap},
         year = 2003,
        month = mar,
       volume = {400},
        pages = {95-118},
          doi = {10.1051/0004-6361:20021861},
       adsurl = {https://ui.adsabs.harvard.edu/abs/2003A&A...400...95N}
}

@ARTICLE{Graham2007MN,
       author = {{Graham}, Alister W.},
        title = "{The black hole mass - spheroid luminosity relation}",
      journal = {\mnras},
         year = 2007,
        month = aug,
       volume = {379},
       number = {2},
        pages = {711-722},
          doi = {10.1111/j.1365-2966.2007.11950.x},
archivePrefix = {arXiv},
       eprint = {0705.0618},
 primaryClass = {astro-ph},
       adsurl = {https://ui.adsabs.harvard.edu/abs/2007MNRAS.379..711G}
}

@ARTICLE{Brinkmann1997,
       author = {{Brinkmann}, W. and {Siebert}, J. and {Feigelson}, E.~D. and {Kollgaard}, R.~I. and {Laurent-Muehleisen}, S.~A. and {Reich}, W. and {F{\"u}rst}, E. and {Reich}, P. and {Voges}, W. and {Truemper}, J. and {McMahon}, R.},
        title = "{Radio-loud active galaxies in the northern ROSAT All-Sky Survey. II. Multi-frequency properties of unidentified sources.}",
      journal = {\aap},
         year = 1997,
        month = jul,
       volume = {323},
        pages = {739-748},
       adsurl = {https://ui.adsabs.harvard.edu/abs/1997A&A...323..739B}
}

@ARTICLE{Wu2009RAA,
       author = {{Wu}, Zhong-Zu and {Gu}, Min-Feng and {Jiang}, Dong-Rong},
        title = "{The debeamed luminosity, sychrotron peak frequency and black hole mass of BL Lac objects}",
      journal = {RAA},
         year = 2009,
        month = feb,
       volume = {9},
       number = {2},
        pages = {168-178},
          doi = {10.1088/1674-4527/9/2/006},
archivePrefix = {arXiv},
       eprint = {0804.1180},
 primaryClass = {astro-ph},
       adsurl = {https://ui.adsabs.harvard.edu/abs/2009RAA.....9..168W}
}

@ARTICLE{Fan2004ApJ,
       author = {{Fan}, Zhong-Hui and {Cao}, Xinwu},
        title = "{Black Hole Masses and Doppler Factors of Gamma-Ray Active Galactic Nuclei}",
      journal = {\apj},
         year = 2004,
        month = feb,
       volume = {602},
       number = {1},
        pages = {103-110},
          doi = {10.1086/380902},
archivePrefix = {arXiv},
       eprint = {astro-ph/0310590},
 primaryClass = {astro-ph},
       adsurl = {https://ui.adsabs.harvard.edu/abs/2004ApJ...602..103F}
}

@ARTICLE{O-Dowd2005ApJ,
       author = {{O'Dowd}, Matthew and {Urry}, C. Megan},
        title = "{Host Galaxy Evolution in Radio-Loud Active Galactic Nuclei}",
      journal = {\apj},
         year = 2005,
        month = jul,
       volume = {627},
       number = {1},
        pages = {97-124},
          doi = {10.1086/426705},
archivePrefix = {arXiv},
       eprint = {astro-ph/0411099},
 primaryClass = {astro-ph},
       adsurl = {https://ui.adsabs.harvard.edu/abs/2005ApJ...627...97O}
}

@ARTICLE{Plotkin2011MN,
       author = {{Plotkin}, R.~M. and {Markoff}, S. and {Trager}, S.~C. and {Anderson}, S.~F.},
        title = "{Dynamical black hole masses of BL Lac objects from the Sloan Digital Sky Survey}",
      journal = {\mnras},
         year = 2011,
        month = may,
       volume = {413},
       number = {2},
        pages = {805-812},
          doi = {10.1111/j.1365-2966.2010.18172.x},
archivePrefix = {arXiv},
       eprint = {1012.1601},
 primaryClass = {astro-ph.CO},
       adsurl = {https://ui.adsabs.harvard.edu/abs/2011MNRAS.413..805P}
}
\bibliographystyle{aasjournalv7}



\end{document}